\documentclass{biometrika}

\usepackage{amsmath}

\DeclareMathOperator*{\argmin}{arg\,min}
\usepackage{amsfonts}
\usepackage{hyperref}
\usepackage{caption}
\usepackage{graphics}
\graphicspath{{Fig/}}

\usepackage{times}
\usepackage{bm}
\usepackage{natbib}
\usepackage{xcolor}

\usepackage{newtxtext}
\usepackage[subscriptcorrection]{newtxmath}
\usepackage{xr}
\usepackage[plain,noend]{algorithm2e}

\makeatletter
\renewcommand{\algocf@captiontext}[2]{#1\algocf@typo. \AlCapFnt{}#2} 
\def\@algocf@capt@plain{top}
\renewcommand{\algocf@makecaption}[2]{%
  \addtolength{\hsize}{\algomargin}%
  \sbox\@tempboxa{\algocf@captiontext{#1}{#2}}%
  \ifdim\wd\@tempboxa >\hsize
    \hskip .5\algomargin%
    \parbox[t]{\hsize}{\algocf@captiontext{#1}{#2}}
  \else%
    \global\@minipagefalse%
    \hbox to\hsize{\box\@tempboxa}
  \fi%
  \addtolength{\hsize}{-\algomargin}%
}
\makeatother

\begin{document}

\jname{Submitted to Biometrika}

\markboth{JIAZHEN Xu et~al.}{Quantifying Periodicity in Non-Euclidean Object-Valued Time Series}

\title{Quantifying Periodicity in Non-Euclidean Object-Valued Time Series}

\author{JIAZHEN XU}
\affil{Department of Actuarial Studies and Business Analytics, Macquarie University, Sydney, 2113, Australia
\email{jiazhen.xu@mq.edu.au}}

\author{ANDREW T.A. WOOD}
\affil{Research School of Finance, Actuarial Studies and Statistics, Australian National University, Canberra ACT 2600, Australia \email{andrew.wood@anu.edu.au}}

\author{TAO ZOU}
\affil{Research School of Finance, Actuarial Studies and Statistics, Australian National University, Canberra ACT 2600, Australia
\email{tao.zou@anu.edu.au}}

\maketitle

\begin{abstract}
Non-Euclidean object-valued time series are playing a growing role in modern data analysis, and periodicity is a fundamental characteristic of many time series. However, quantifying periodicity in general non-Euclidean random objects remains largely unexplored. In this work, we introduce a novel nonparametric framework for quantifying periodicity in random objects within a general metric space that lacks Euclidean structure. Our approach formulates periodicity estimation as a model selection problem and provides methodologies for period estimation, data-driven tuning parameter selection, and periodic component extraction. Our theoretical contributions include deriving uniform convergence rates for general Fr\'{e}chet regression under temporal dependence, establishing the consistency of period estimation without relying on linearity properties used in the literature for Euclidean data, providing theoretical support for data-driven tuning parameter selection, and deriving uniform convergence results for periodic component estimation. Through extensive simulation studies covering three distinct types of time-varying random objects such as compositional data, networks, and functional data, we showcase the excellent accuracy achieved by our approach in periodicity quantification. Finally, we apply our method to various real datasets, including compositional data arising in U.S. electricity generation, New York City transportation networks, and Germany’s water consumption curves, highlighting its practical relevance in identifying and quantifying meaningful periodic patterns.
\end{abstract}


\section{Introduction}

Advances in data collection technologies have enabled the increasingly widespread acquisition of random objects (\citealt{marron2021object}). Such objects arise in diverse forms, including networks, phylogenetic trees, probability distributions, covariance matrices, compositional data, and spherical data (\citealt{kolar2010estimating, zhou2022network, scealy2023score}). When these objects are repeatedly observed over time, they give rise to object-valued time series (\citealt{dubey2021modeling,zhu2024spherical}). A fundamental challenge in analyzing such data is that the underlying objects often belong to general metric spaces rather than Euclidean spaces. These spaces typically lack the linear structure and vector-space operations that underpin conventional statistical methodology, making standard analytical techniques difficult or inappropriate to apply directly.

Periodic patterns are frequently observed in object-valued time series. For instance, \cite{hormann2018testing} study periodicity in functional time series, identifying weekly cycles in daily air pollution curves caused by weekday-dependent traffic patterns in Graz, Austria. \cite{xu2025change} observe periodic behaviour of time-varying transportation networks in New York City. Dynamic transportation networks observed on an hourly basis are influenced by rush hour patterns, resulting in periodicity with period being assumed to be twenty-four hours.

Further examples of periodic random objects arise in applications such as electricity generation and social communication networks. In electricity generation, monthly observations can be represented by the percentage contributions of different energy sources to total net generation, and a 12-month period for the observations can be assumed. Social communication networks provide another example, as their graph structures evolve over time, with patterns of connections and communication activity often exhibiting recurring temporal fluctuations.

\begin{figure}[!htbp]
\centering
\includegraphics[width=1\linewidth]{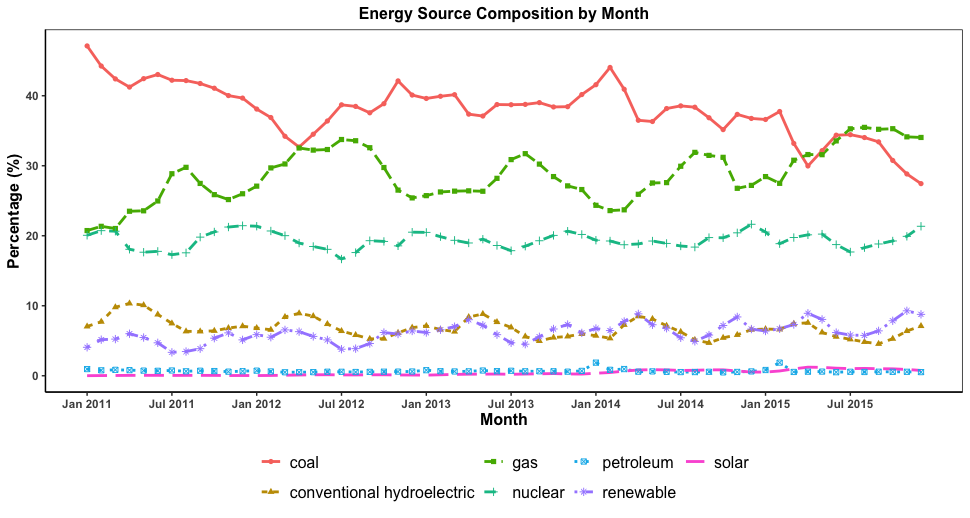}

\caption{The timeline of U.S. electricity generation compositions from January 2011 to December 2015. Each line represents the percentage of the corresponding energy resource categories. \label{fig::electricity generartion sample}}
\end{figure}

All the examples above exhibit periodic behaviour, but the specific periodicity, such as the length of the period, is often an assumption rather than a certainty. For example, Fig.~\ref{fig::electricity generartion sample} shows the timeline of U.S. electricity generation compositions from January 2011 to December 2015, considering 7 major resource categories. It is hard to directly discern a clear periodic pattern and thus there is a need for the period to be rigorously quantified using suitable statistical tools.

Notably, periodicity is a fundamental characteristic of time series, and ignoring it can have negative effects. For instance, prediction intervals tend to be more precise when the periodic pattern of the data is accounted for, whereas neglecting periodicity often results in overly wide intervals. Furthermore, \cite{guo2020nonparametric} and \cite{xu2025change} demonstrate that neglecting periodicity can degrade the performance of change point detection, as periodic patterns may blur the actual change points. These examples also underscore the importance of quantifying periodicity in data analysis.

So far, numerous methods on periodicity quantification have mainly focused on univariate real-valued time series. These techniques fall into two broad categories: parametric approaches, which model periodic patterns using trigonometric regressions (\citealt{walker1971estimation, rice1988frequency, quinn1991estimating}), and nonparametric methods, which frame periodicity estimation as a model selection problem (\citealt{hall2000nonparametric, sun2012nonparametric, vogt2014nonparametric}). However, despite a rich literature on periodicity estimation for Euclidean data, quantifying periodicity in general non-Euclidean object-valued time series remains largely unexplored. Although the extension of trigonometric regression methods for functional data, a special case of object data, has been studied in \cite{hormann2018testing}, their approach, like other parametric methods for periodic Euclidean data, relies on the linear structure of the data space and is inapplicable to general metric spaces without any local or global linear structure.

Motivated by these challenges, this paper introduces a novel nonparametric framework for periodicity quantification for object-valued time series. Our framework consists of two main components: (i) estimation of the unknown period and (ii) quantification of the periodic component based on the estimated period. For period estimation, we adapt the idea of \cite{vogt2014nonparametric} from Euclidean data to non-Euclidean data where we formulate the period estimation problem as one of model selection. The key step relies on constructing a suitable class of models designed to produce a high goodness-of-fit if and only if the candidate period is a positive integer multiple of the true but unknown period. Moreover, once a consistent estimate of the period is obtained, the constructed model is then used to estimate the periodic component. Our proposed approach represents a significant advancement in periodicity quantification for random objects. To the best of our knowledge, this is the first general framework proposed for this problem, extending beyond functional data to a wide range of non-Euclidean random objects. 

In addition to these methodological contributions, we establish rigorous theoretical support for the proposed framework. Specifically, the absence of a linear structure in general metric spaces necessitates a significant departure from the proof techniques commonly used in nonparametric period estimation (e.g., \cite{vogt2014nonparametric} and \cite{wang2022robust}), which typically rely on an explicit additive error structure. Furthermore, we develop information criteria for data-driven tuning parameter selection to guarantee the consistency of the period estimator, and we provide general theoretical guidance for constructing these criteria. This addresses a notable gap in the literature, as such data-driven selection mechanisms were not considered in \cite{vogt2014nonparametric} and \cite{wang2022robust}, even for Euclidean data. The core of our theoretical development rests on establishing the uniform convergence rate of a generic Fr\'{e}chet regression estimator under a temporal dependence setting, with a special case being the global Fr\'{e}chet regression proposed by \cite{petersen2019frechet}. While the existing literature on Fr\'{e}chet regression predominantly relies on the assumption of independent observations; see, e.g. \cite{petersen2019frechet,zhou2022network,hong2026uniform}, this constraint is inherently restrictive for time series analysis. By accommodating temporal dependence, our work overcomes this limitation. Establishing uniform convergence rates in this dependent, non-linear regime requires novel technical arguments to control the fluctuations of certain empirical processes on metric spaces, and significantly broadens the applicability of Fr\'{e}chet regression to the temporal setting.

Extensive simulations on periodic random objects, including spherical data, networks, and distributional data, showcase the superior accuracy of our method in both period estimation and periodic component quantification. We further validate its effectiveness through applications to three real-world datasets: (i) monthly U.S. electricity generation compositions, (ii) hourly transportation networks from the New York City Citi Bike system, and (iii) daily water consumption curves in Germany. In each case, our method provides interpretable and meaningful insights into the periodic structure of the data.


\section{Methodology}\label{sect::methodology}

In this section, we first present the problem setup for periodic object-valued time series. We introduce nonparametric methodologies for period estimation, tuning parameter selection, and periodic component estimation in \S\ref{subsection::period estimation} and \ref{subsection::periodic component estimation}.

\subsection{Preliminaries}\label{subsection::problem setup}

Suppose we have a object-valued time series $\{Y_1,Y_2,\ldots,Y_T\}$ where $Y_t$ locates within a totally bounded metric space $\Omega$ with metric $d$ for each $t=1,2,\ldots,T$, and that the time series has a periodic pattern with the true smallest period being a positive integer $\theta_0$. For generic probability measures $P^{(l)}$ on $(\Omega,d)$ with $l\in\{1,\ldots,\theta_0\}$, $\{Y_t\}_{t=1}^T$ is said to follow a $\theta_0$-periodic distribution if there exists a positive integer $\theta_0$ such that, for any integer $j\in\{1,\ldots,\theta_0\}$, (i) $Y_j\sim P^{(j)}$ and (ii) $Y_j,Y_{j+\theta_0},Y_{j+2\theta_0},\ldots,Y_{j+k\theta_0}$ have the same distribution for integer $k\geq 0$ satisfying $j+k\theta_0\leq T$. Note that here we only require the stationarity of $\{Y_1,Y_2,\ldots,Y_T\}$ with respect to the marginal distributions $\{P^{(l)}\}_{l=1}^{\theta_0}$. 

For such periodic random objects $Y_1,Y_2,\ldots,Y_T$ with period $\theta_0$, the corresponding population Fr\'{e}chet mean (\citealt{frechet1948elements}) can be defined by
\begin{align}\label{formula::periodic component}
     m(r(t,\theta_0)) = \argmin_{\nu\in\Omega} E \{ d^2(Y_t,\nu) \},    
\end{align}
where $r(t,\theta_0)=t+\theta_0-\theta_0\lfloor (t+\theta_0-1)/\theta_0 \rfloor$ with $\lfloor\cdot\rfloor$ being the floor function, and $m(\cdot)\in\Omega$ is the deterministic periodic component with period $\theta_0$ and domain being the set of positive integers such that $m(l)=m(l+\theta_0)=m(l+2\theta_0)=\cdots$ for $l\in\{1,\ldots,\theta_0\}$.

In this framework, we do not assume any specific parametric structure of the periodic component $m(\cdot)$, and $m(\cdot)$ is treated as a sequence rather than a $\Omega$-valued function. Our main focus is the equidistant design which is the most common situation, as seen in the real data examples considered.

\subsection{Period Estimation for Random Objects}\label{subsection::period estimation}

A common way to estimate the period for real-valued time series is to formulate the period estimation problem as one of model selection, first proposed by \cite{vogt2014nonparametric}. Inspired by this idea, consider a set of candidate periods $\theta \in \{1, 2, \ldots, \Theta_T\}$, where $\Theta_T$ represents the maximum allowable period length. Here, the upper bound $\Theta_T$ is allowed to grow with the sample size $T$ at a suitable rate, see Theorem \ref{thm::period estimation consistency}. For any given candidate period $\theta$, consider a sequence of fitted models $\widehat{m}_t(\theta, T)$ for $t=1,2,\ldots,T$,with the corresponding residual sum of squares (RSS) given by ${\rm RSS}(\theta) = \sum_{t=1}^T d^2(Y_t , \widehat{m}_t(\theta, T))$. We informally describe the conditions under which these models $\{\widehat{m}_t(\theta, T) \}_{t=1}^T$ are suitable for period estimation in Definition~\ref{def::suitable model}.

\begin{definition}\label{def::suitable model}
A model sequence $\{ \widehat{m}_t(\theta, T) \}_{t=1}^T$ is suitable for period estimation if, with high probability, ${\rm RSS}(\theta) = \sum_{t=1}^T d^2(Y_t , \widehat{m}_t(\theta, T))$ satisfies the following three conditions:
\begin{itemize}
\item[1.] Correct Specification. At the true period $\theta = \theta_0$, ${\rm RSS}(\theta)$ is sufficiently small.
\item[2.] Overfitting. For any integer multiple $\theta = k\theta_0$ where $k \geq 2$ is an integer, the model overfits the data, yielding ${\rm RSS}(\theta) < {\rm RSS}(\theta_0)$.
\item[3.] Misspecification. For any $\theta$ that is not an integer multiple of $\theta_0$, the model is misspecified, resulting in ${\rm RSS}(\theta) > {\rm RSS}(\theta_0)$.
\end{itemize}
\end{definition}

The key to the period estimation is the construction of the suitable model $\widehat{m}_t(\theta, T)$ with required properties in Definition~\ref{def::suitable model}. However, in the literature for the real-valued time series, see, \cite{vogt2014nonparametric} and \cite{wang2022robust}, the suitable model are built based on the linear regression which is not applicable to non-Euclidean object-valued time series. To fill this gap, we construct models in the form of
\begin{align}\label{empirical conditional mean formula}
    \widehat{m}_t(\theta, T) = \argmin_{\nu\in\Omega} \widehat{M}_t(\nu;\theta,T),~~\widehat{M}_t(\nu;\theta,T)= \{n(t,\theta)\}^{-1}\sum_{i=1}^{n(t,\theta)} d^2\left(Y_{r(t,\theta)+(i-1)\theta},\nu\right)
\end{align}
where $n(t,\theta)=M(\theta)+\mathbb{I}(r(t,\theta)\leq r(T,\theta))$ with $M(\theta) = \lfloor T/\theta\rfloor$ for any $\theta$ such that $1\leq \theta \leq \Theta_T$
for $t\in\{1,2,\ldots,T\}$. Here, $\mathbb{I}(\cdot)$ is the indicator function and the constructed model $\widehat{m}_t(\theta, T)$ is the empirical Fr\'{e}chet mean of a specified subset of observations from $\{Y_t\}_{t=1}^T$, whose population counterpart is 
\begin{align}\label{conditional mean formula}
    m_t(\theta, T) = \argmin_{\nu\in\Omega} M_t(\nu; \theta, T),~~M_t(\nu; \theta, T)= \{n(t,\theta)\}^{-1}\sum_{i=1}^{n(t,\theta)} E\left\{ d^2\left(Y_{r(t,\theta)+(i-1)\theta},\nu\right)\right\}.
\end{align}
By our construction, $m_t(\theta, T) = m(r(t, \theta_0))$ as defined in (\ref{formula::periodic component}) if and only if $\theta = k\theta_0$ for any positive integer $k$. This population-level equivalence is what ensures the estimator $\widehat{m}_t(\theta, T)$ satisfies the properties outlined in Definition~\ref{def::suitable model}. To illustrate the overfitting property, consider the extreme case where $T = K\theta_0$ for some integer $K > 1$ and the candidate period spans the entire time series $\theta=T$. Under this situation, the estimator perfectly interpolates the data, yielding $\widehat{m}_t(\theta, T) = Y_t$ for all $t = 1,2, \ldots, T$, which drives the RSS to its minimum of zero.

To account for the overfitting at higher multiples of the true period in Definition~\ref{def::suitable model}, we define a penalized RSS as 
\begin{equation}\label{formula::loss function}
\mathcal{L}(\theta,\lambda_T) = {\rm RSS}(\theta) + \lambda_T \theta
\end{equation}
The estimated period $\widehat{\theta}_{\lambda_T}$ is then the candidate that minimises this objective as
\begin{equation}\label{formula::period estimator}
\widehat{\theta}_{\lambda_T} = \argmin_{1\leq \theta \leq \Theta_T} \mathcal{L}(\theta,\lambda_T).
\end{equation}
The tuning parameter $\lambda_T$ is designed to diverge to infinity at an appropriate rate to ensure that the penalty $\lambda_T \theta$ effectively suppresses the overfitting situation; see Theorem \ref{thm::period estimation consistency} for the formal consistency requirements. Note that the penalty term $\lambda_T\theta$ in (\ref{formula::loss function}) can be viewed as an $\ell_0$-penalty and the period estimation problem is now formulated as a model selection problem where for each candidate $\theta\in\{1,\ldots,\Theta_T\}$, we construct a model of the form (\ref{empirical conditional mean formula}). The goal is then to identify the correct model among these candidates.

\begin{figure}[!htbp]
\centering
\includegraphics[width=1\linewidth]{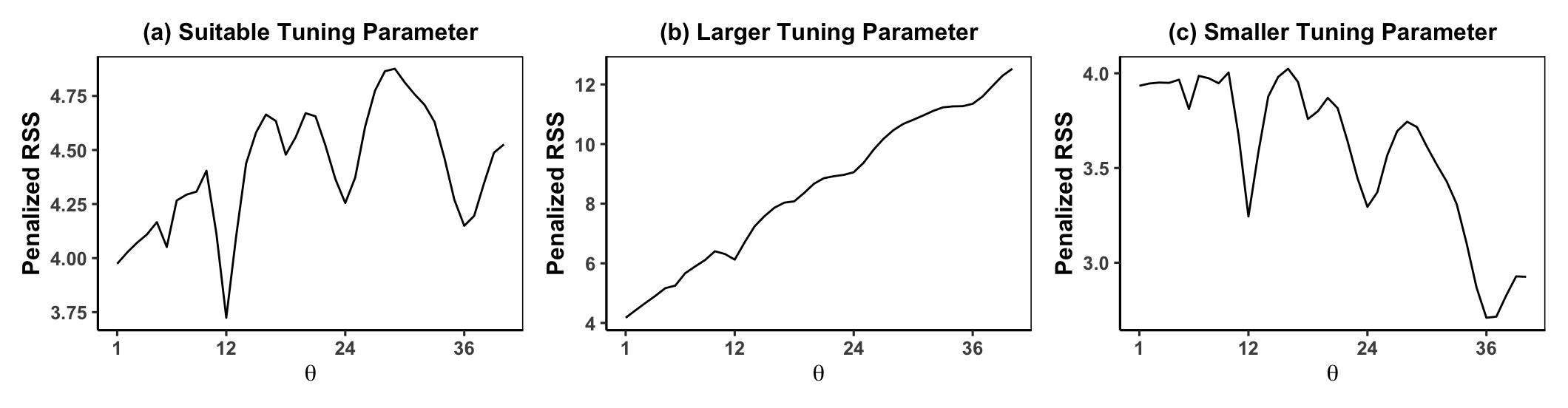}

\caption{ Plots of the penalized RSS for U.S. electricity generation compositions from January 2011 to December 2015 considering different choice of the tuning parameter. In particular, a suitable choice of $\lambda_T=\widehat{\lambda}$ with $\widehat{\lambda}$ selected via (\ref{formula::lambda selection})} in (a), $\lambda_T=5\widehat{\lambda}$ in (b), and $\lambda_T=\widehat{\lambda}/5$ in (c).    \label{fig::toy example composition loss}
\end{figure}

Notably, the visualization of random objects remains an underdeveloped area in the literature, and no existing visualization tool effectively reveals the presence of periodicity in random objects. Our proposed penalized RSS in (\ref{formula::loss function}) provides a powerful tool to address this gap. For example, in Fig.~\ref{fig::toy example composition loss}(a), consider the U.S. electricity generation compositions from January 2011 to December 2015, a pronounced drop in the penalized RSS at a specific value, together with drops at its positive integer multiples, strongly suggests the presence of a periodic pattern in the object data. Figure~\ref{fig::toy example composition loss}(a) indicates that the U.S. electricity generation compositions have a period of 12 while the periodic pattern cannot be directly observed in the line plot of each elements of the compositions as shown in Fig.~\ref{fig::electricity generartion sample}.

In (\ref{formula::loss function}), the tuning parameter $\lambda_T$ plays a crucial role in estimation. As one can see in Figs.~\ref{fig::toy example composition loss}(b) and \ref{fig::toy example composition loss}(c), a tuning parameter that is too small can lead to overfitting, while one that is too large can cause misspecification, resulting in inaccurate estimation of the true period $\theta_0$. Motivated by the Bayesian information criterion (\citealt{schwarz1978estimating}), we develop a data-driven information criterion for selecting the tuning parameter. Notably, existing nonparametric methods for Euclidean data, such as those proposed by \cite{vogt2014nonparametric} and \cite{wang2022robust}, do not incorporate a data-driven approach for tuning parameter selection. This constitutes one of the contributions of our work.

Here, we consider a general regularization function $g(T)$, such that an information criterion of the form
\begin{align} \label{eq::BIC}
 {\rm IC}_\lambda = \log \{ {\rm RSS}(\widehat{\theta}_\lambda) / T \} + \widehat{\theta}_\lambda g(T),~~{\rm or}~~
    {\rm IC}_\lambda = {\rm RSS}(\widehat{\theta}_\lambda) / T  + \widehat{\theta}_\lambda g(T),
\end{align}
can consistently estimate the true period $\theta_0$. 
We aim to develop a general theoretical framework for the information criterion (\ref{eq::BIC}) and the conditions for constructing the general regularization function $g(T)$ in (\ref{eq::BIC}) are given in Theorem \ref{thm::IC} below. An analogous framework can be found in the area of factor models; see, e.g. \cite{bai2002determining}.
Using (\ref{eq::BIC}), the tuning parameter $\lambda$ may be selected as 
\begin{align}\label{formula::lambda selection}
    \widehat{\lambda}=\argmin_\lambda {\rm IC}_\lambda.
\end{align}
Based on the information criteria in (\ref{formula::lambda selection}), the final period estimator is given by $\widehat{\theta}_{\widehat{\lambda}}$, obtained from (\ref{formula::period estimator}) by setting $\lambda_T=\widehat{\lambda}$.

\subsection{Periodic Component Estimation for Random Objects}\label{subsection::periodic component estimation}

Write $\widehat{\theta}_{\widehat{\lambda}}$  for the estimator of the true but unknown period $\theta_0$. Let $\mathbb{N}^+$ be the set of positive integers. Then an estimator of the periodic sequence $\{m(t)\}_{t\in\mathbb{N}^+}$ given by (\ref{formula::periodic component}) is
\begin{align}\label{formula::periodic component estimator}
    \widehat{m}(l)=\widehat{m}_l\left(\widehat{\theta}_{\widehat{\lambda}},T\right),
\end{align}
and $\widehat{m}(l+k\widehat{\theta}_{\widehat{\lambda}}) = \widehat{m}(l)$ for $l=1,2,\ldots,\widehat{\theta}_{\widehat{\lambda}}$ and all $k=0,1,2,\ldots$. This construction ensures that $\{\widehat{m}(t)\}_{t=1,\ldots,T}$ is a periodic sequence with period $\widehat{\theta}_{\widehat{\lambda}}$.

Notice that the periodic component estimator is a statistic with the additional estimated parameter, $\widehat{\theta}_{\widehat{\lambda}}$. This will pause some challenges in establishing theoretical results as this additional estimated parameter involves additional randomness. To deal with the statistic with estimated parameters, the classical way and the method adopted in the proof is to first introduce an oracle estimator $\widetilde{m}(l)$ in the case where the true period $\theta_0$ is assumed to be known. We then evaluate the asymptotic behaviours of $d(\widehat{m}(l),\widetilde{m}(l))$ and $d(\widetilde{m}(l),m(r(l,\theta_0)))$, respectively.

\section{Theoretical Results}\label{sect::theory}

Before presenting the main theoretical results, we first define the $\alpha$-mixing coefficient for the object-valued time series $\{Y_t\}_{t=1}^T$. For random objects $\{Y_t\}_{t\in\mathbb{Z}}$, denote the $\sigma$-field generated by $\{Y_{t_1},Y_{t_1+1},\ldots,Y_{t_2}\}$ with $-\infty\leq t_1\leq t_2 \leq \infty$ as $\mathcal{F}_{t_1}^{t_2}$. The sequence $\{Y_t\}_{t\in\mathbb{Z}}$ is said to be $\alpha$-mixing if the $\alpha$-mixing coefficients $\alpha(n) \to 0$ as $n \to \infty$, where $\alpha(n)=\sup_{\mathcal{A}\in\mathcal{F}_{-\infty}^0,\mathcal{B}\in\mathcal{F}_n^{\infty}}\left| {\rm pr}(\mathcal{A}\cap \mathcal{B})-{\rm pr}(\mathcal{A}){\rm pr}(\mathcal{B}) \right|.$

The core of our theoretical development rests on establishing the uniform convergence rate of a generic Fr\'{e}chet regression estimator. For a bounded predictor space $\mathcal{X}$ with the metric $d_{\mathcal{X}}$, consider generic fixed weights $w_1(x),w_2(x),\cdots,w_T(x)$, and define the Fr\'{e}chet regression estimator with fixed weights as 
\begin{align*}
    \varkappa (x)=\argmin_{\nu\in\Omega}\Xi(x,\nu),~~\Xi(x,\nu)=T^{-1}\sum_{t=1}^T w_t(x) E\{ d^2(Y_t,\nu) \},
\end{align*}
and 
\begin{align}\label{formula::generic global Frechet regression}
    \widehat{\varkappa}(x)=\argmin_{\nu\in\Omega}\widehat{\Xi}(x,\nu),~~\widehat{\Xi}(x,\nu)=T^{-1}\sum_{t=1}^T w_t(x)  d^2(Y_t,\nu).
\end{align}
See Models~3.1--3.5 in \cite{hong2026uniform} for a summary of different types of Fr\'{e}chet regression models in the form of (\ref{formula::generic global Frechet regression}).

The following technical assumptions are necessary to derive the uniform convergence rate of $\sup_{x\in\mathcal{X}}d(\widehat{\varkappa}(x),\varkappa (x))$.

\begin{assumption}\label{assumption 1}
    $\{Y_t\}_{t\in\mathbb{Z}}$ is $\alpha$-mixing such that $\alpha(n) \leq \exp(-2C n)$ for some positive constant $C$.
\end{assumption}

\begin{assumption}\label{assumption 2}
(i) $\Omega$ has a totally bounded separable and convex subset $\mathcal{C}$ with a metric $d_a$ such that $m_t(x),\widehat{\varkappa}(x)\in \mathcal{C}$ with probability tending to one as $T \rightarrow \infty$. The convex set $\mathcal{C}$ means for every $x,y\in\mathcal{C}$, there exists $z\in\mathcal{C}$ such that $d_a(x,z)+d_a(z,y)=d_a(x,y)$. (ii) Suppose there exists $\widetilde{\varkappa}\in \mathcal{C}$ and $\widetilde{\varkappa}\notin\{\varkappa(x),\widehat{\varkappa}(x)\}$ such that $d_a(\varkappa(x),\widetilde{\varkappa})+d_a(\widetilde{\varkappa},\widehat{\varkappa}(x))=d_a(\varkappa(x),\widehat{\varkappa}(x))$ and 
\[
\widehat{\Xi}(\widetilde{\varkappa},x) \leq \delta(x) \widehat{\Xi}(\widehat{\varkappa}(x),x) + \{1-\delta(x)\} \widehat{\Xi}(\varkappa(x),x)
\]
holds for all $x\in\mathcal{X}$ where $\delta(x)=d_a(\varkappa(x),\widetilde{\varkappa})/d_a(\varkappa(x),\widehat{\varkappa}(x))$.
\end{assumption}

\begin{assumption}\label{assumption 3}
    For any $x\in\mathcal{X}$, each barycenter $\varkappa(x)$ exists, and
    \[
    \inf_{x\in\mathcal{X}}\inf_{\nu\in\Omega:d(\nu,\varkappa(x))>\gamma} \Xi(x,\nu)-\Xi(x,\varkappa(x))>0
    \]
    for all $\gamma>0$.
\end{assumption}

\begin{assumption}\label{assumption 4}
    There exist $\zeta>0$, $C>0$, and $\beta>1$ such that,
    \[
    \inf_{x\in\mathcal{X}}\inf_{d(\nu,\varkappa(x))<\zeta}\left\{\Xi(x, \nu) - \Xi(x, \varkappa(x)) - C d^\beta(\varkappa(x), \nu)\right\}>0.
    \]
\end{assumption}

\begin{assumption}\label{assumption 5}
    For $I(\zeta)=\sup_{x\in\mathcal{X}}\int_0^1\sqrt{1+\log N(\delta\zeta,B_\zeta(\varkappa(x)),d)}d\delta$, it holds that $I(\zeta)=O(1)$ as $\zeta\to0$ for each $x\in\mathcal{X}$, where $B_\zeta(\varkappa(x))=\{\nu\in\Omega:d(\nu,\varkappa(x))<\zeta \}$ is the $\zeta$-ball around $\varkappa(x)$ and $N(\gamma,B_\zeta(\varkappa(x)),d)$ is the covering number.
\end{assumption}

Assumption~\ref{assumption 1} assumes that the sequence $\{Y_t\}_{t\in\mathbb{Z}}$ is geometrically $\alpha$-mixing, a condition necessary for applying the Bernstein inequality in Theorem 2 of \cite{merlevede2009bernstein}. Assumption~\ref{assumption 2}, adapted from \cite{hong2026uniform}, consists of two parts, where part (i) imposes a geometric structural condition on a subset of $\Omega$ that guarantees the existence of an intermediate point $\widetilde{\varkappa}$ required in part (ii); and part (ii) serves as a slightly weaker form of the local convexity requirement on the empirical cost function $\widehat{\Xi}(\nu,x)$ within the neighborhood of $\varkappa(x)$. If $\Omega$ is a convex subset of a Euclidean space and both $d$ and $d_a$ are the induced Euclidean distances, Assumption~\ref{assumption 2} simplifies to the standard local convexity condition found in classical $M$-estimation theory. For general geodesic spaces, this assumption formalizes the local geodesic convexity of the empirical cost function. Various metric spaces satisfy this condition, including Riemannian manifolds and Hadamard spaces, see Propositions 2--6 in \cite{hong2026uniform}. Assumption~\ref{assumption 3} establishes an identifiability condition that ensures both the existence and uniqueness of the population conditional barycenter for any $x\in\mathcal{X}$, see \cite{sturm2003probability} and \cite{ahidar2020convergence} for related discussions. Together, Assumptions~\ref{assumption 2} and \ref{assumption 3} allow us to utilize Theorem 2 of \cite{hong2026uniform} to establish the uniform consistency $\sup_{x\in\mathcal{X}}d(\widehat{\varkappa}(x),\varkappa (x)=o_p(1)$, a fundamental prerequisite for deriving the convergence rate. Finally, Assumptions~\ref{assumption 4} and \ref{assumption 5} dictate this uniform convergence rate. Assumption~\ref{assumption 4} quantifies the uniform local curvature of the cost function near its minimum via a parameter $\beta$, while Assumption~\ref{assumption 5} bounds the covering number, defined e.g. in \cite{vanderVaartWellner2023}, of the object metric space. Assumptions~\ref{assumption 3}--\ref{assumption 5} align with those utilized by \cite{petersen2019frechet} for the uniform convergence rate of conditional barycenters. Object spaces satisfying Assumptions~\ref{assumption 2}--\ref{assumption 5} with $\beta=2$ in Assumption~\ref{assumption 4} include graph Laplacian matrices equipped with the Frobenius metric, univariate probability distributions under the 2-Wasserstein metric, and spherical data with the geodesic distance (\citealt{petersen2019frechet,hong2026uniform}).

The theorem below provides the uniform convergence rate of $\sup_{x\in\mathcal{X}}d(\widehat{\varkappa}(x),\varkappa(x))$. 

\begin{theorem}\label{thm::uniform conv}
    Under Assumptions~\ref{assumption 1}--\ref{assumption 5}, suppose $w_{\rm max}=\sup_{t\in\{1,2,\cdots,T\}}\sup_{x\in\mathcal{X}}|w_t(x)|<\infty$, $|w_t(x)-w_t(x^\prime)|\leq L_w d_{\mathcal{X}} (x,x^\prime)$ for some constant $L_w$, and the covering number $N(r, \mathcal{X}, d_{\mathcal{X}}) = O(e(T))$ for any $r>0$ where $\{\log e(T) \}/T=o(1)$. Then
    \[
    \sup_{x\in\mathcal{X}}d(\widehat{\varkappa}(x),\varkappa(x)) =O_p\left(\left(\frac{T}{\log e(T)}\right)^{-1/\{2(\beta^\prime-1)\}}\right)
    \]
    for any $\beta^\prime>\beta$ with $\beta$ given in Assumption~\ref{assumption 4}.
\end{theorem}

\begin{remark}
    Theorem~\ref{thm::uniform conv} relaxes three commonly assumed assumptions in related literature for Fr\'{e}chet regression. First, it accommodates $\alpha$-mixing observations $Y_1,Y_2, \ldots, Y_T$, thereby relaxing the common assumption of  independence. Second, it allows the covering number of the predictor space $\mathcal{X}$ diverges as $T\to\infty$, relaxing the  common assumption of total boundedness. Third, it does not assume a high-level uniform assumption to achieve uniform asymptotic equicontinuity in probability of the Fr\'{e}chet regression estimator; see, e.g., the discussion in Section~2.1 in \cite{hong2026uniform}. In contrast, Theorem~\ref{thm::uniform conv} is built on the analytical condition of the local convexity of the cost function in Assumption~\ref{assumption 2} that bypasses the need to verify uniform equicontinuity.  
\end{remark}
\begin{remark}
    If the predictor space $\mathcal{X}$ is further assumed to be totally bounded, Theorem~\ref{thm::uniform conv} will give $\sup_{x\in\mathcal{X}}d(\widehat{\varkappa}(x),\varkappa(x)) =O_p\left(T^{-1/\{2(\beta^\prime-1)\}}\right)$, which is the same rate as in Theorem 2 in \cite{petersen2019frechet} for global Fr\'{e}chet regression.  
\end{remark}

The temporal dependence introduces substantial technical challenges, requiring standard empirical process tools to be adapted for the dependent setting, while the diverging covering number introduces challenges in entropy calculation. The proof of Theorem~\ref{thm::uniform conv} is structured in two main parts. First, we establish uniform consistency $\sup_{x\in\mathcal{X}}d(\widehat{\varkappa}(x),\varkappa(x))=o_p(1)$, by applying Theorem 2 of \cite{hong2026uniform} and verifying the uniform convergence of the empirical cost function. Second, building on this uniform consistency result, we employ a chaining argument to derive the uniform convergence rate. This latter step involves establishing a Bernstein-type inequality based on Theorem 2 of \cite{merlevede2009bernstein}, deriving a maximal inequality utilizing Theorem 2.2.19 of \cite{vanderVaartWellner2023}, and synthesizing peeling techniques with entropy calculation.

\subsection{Theoretical Results for the Period Estimator and Tuning Parameter Selection}

We now apply Theorem~\ref{thm::uniform conv} to establish the theoretical supports for the proposed period estimation method. To achieve this, we need to represent $\widehat{m}_t(\theta,T)$ in a form of $\widehat{\varkappa}(x)$ in (\ref{formula::generic global Frechet regression}) with the required conditions of the fixed weights in Theorem~\ref{thm::uniform conv} being satisfied. For $x=(x_1,x_2,\ldots,x_{\theta})^\top$, let $w_t(x)=x_{r(t,\theta)}$, and thus we can rewrite 
\begin{align}\label{formula::specific Frechet regression}
    \widehat{\varkappa}(x)=\argmin_{\nu\in\Omega}\widehat{\Xi}(x,\nu),~~\widehat{\Xi}(x,\nu)=\{n(t,\theta)\}^{-1}\sum_{t=1}^T w_t(x)  d^2(Y_t,\nu).
\end{align}
The predictor space $\mathcal{X}$ is
\begin{align}\label{curlyX}
    \mathcal{X}=\left \{x \in\mathbb{R}^{\theta}: x=e_{\theta,j}, \, 1 \leq \theta \leq \Theta_T, \, j=1,2,\ldots , T\right \},
\end{align}
where $e_{\theta,t}$ is the $j$th  coordinate vector, i.e., the $\theta \times 1$ vector whose $j$th component is 1 and whose other components are zero, where $j=r(t,\theta)$. This construction gives $\widehat{m}_t(\theta,T)=\widehat{\varkappa}(e_{\theta,t})$. Under this setting, the following proposition provides the specific uniform convergence rate of $\sup_{x\in\mathcal{X}}d(\widehat{\varkappa}(x),\varkappa(x))$.


\begin{proposition}\label{prop::uniform convergence rate partial mean}
    Assume that $\Theta_T\to\infty$ and $(\Theta_T\log\Theta_T)/T \to 0$ as $T\to\infty$. Then for the Fr\'{e}chet regression in (\ref{formula::specific Frechet regression}) with the corresponding population counterpart $\varkappa(x)$, and the predictor space $\mathcal{X}$ given in (\ref{curlyX}), we have, under Assumptions~\ref{assumption 1}--\ref{assumption 5},
    \[
    \sup_{x\in\mathcal{X}}d(\widehat{\varkappa}(x),\varkappa(x)) =O_p\left(\left(\frac{T}{\Theta_T\log \Theta_T}\right)^{-1/\{2(\beta^\prime-1)\}}\right)
    \]
    for any $\beta^\prime>\beta$ with $\beta$ given in Assumption~\ref{assumption 4}.
\end{proposition}

Theorem~\ref{thm::period estimation consistency} given below is built upon Proposition~\ref{prop::uniform convergence rate partial mean} and characterizes the asymptotic behaviour of the period estimator $\widehat{\theta}_\lambda$ given by (\ref{formula::period estimator}). To concisely express the requirement for the tuning parameter, denote $a_T\ll b_T$ or $b_T\gg a_T$ as $a_T=o(b_T)$  for any two sequences $\{a_T\}$ and $\{b_T\}$. 

\begin{theorem}\label{thm::period estimation consistency}
        Assume that $\Theta_T\to\infty$ and $(\Theta_T\log\Theta_T)/T \to 0$ as $T\to\infty$. Select the tuning parameter $\lambda_T$ such that $T^{1-1/(\beta^\prime-1)}\left(\Theta_T\log\Theta_T\right)^{1/(\beta^\prime-1)}\ll \lambda_T \ll T$ for any $\beta^\prime>\beta$ with $\beta$ given in Assumption~\ref{assumption 4}. Then $\widehat{\theta}_{\lambda_T}=\theta_0+o_p(1)$ as $T\to\infty$ under Assumptions~\ref{assumption 1}--\ref{assumption 5}.
\end{theorem}
\begin{remark}
    Theorem \ref{thm::period estimation consistency} is developed by analysing $\mathcal{L}(\theta)$ in two sets $\mathcal{M}_1=\{\theta:\theta =k\theta_0,k=2,\ldots,~{\rm and}~1\leq \theta\leq \Theta_T\}$ and $\mathcal{M}_2=\{\theta:\theta\neq\theta_0,\theta\notin \mathcal{M}_1,~{\rm and}~1\leq \theta\leq \Theta_T\}$. $\mathcal{M}_1$ corresponds to the overfitting situation while $\mathcal{M}_2$ can be interpreted as the misspecification case. When $\Theta_T\to\infty$, the cardinality of either $\mathcal{M}_1$ or $\mathcal{M}_2$ is not finite any more, resulting in the need for the uniform result for $\widehat{m}_t(\theta,T)$ over $\theta\in\{1,2,\ldots,\Theta_T\}$. See the proof of  Theorem~\ref{thm::period estimation consistency} in the Supplementary Material for details. A discussion of a similar situation in tuning parameter selection can be found in \S2.2 of \cite{wang2009shrinkage}.
\end{remark}
\begin{remark}\label{remark::1}
    Metric spaces that satisfy Assumption~\ref{assumption 4} with $\beta=2$ include symmetric positive definite matrices of fixed dimensions with the Frobenius metric, graph Laplacian matrices with the Frobenius metric, univariate probability distributions with the 2-Wasserstein metric, and the unit sphere with the geodesic distance metric (see \cite{petersen2019frechet} and \cite{dubey2021modeling}).
\end{remark}

Theorem \ref{thm::period estimation consistency} establishes the consistency result of the proposed period estimator $\widehat{\theta}_{\lambda_T}$ under a broad class of tuning sequences of $\lambda_T$. Specifically, the result holds for all $\lambda_T$ satisfying $T^{1-1/(\beta^\prime-1)}(\log \Theta_T)^{1/(\beta_2^\prime-1)}\ll \lambda_T \ll T$. This flexible range accommodates a variety of growth rates for the tuning parameter $\lambda_T$. This means, selecting any $\lambda_T$ satisfying this flexible range of condition can lead to estimation consistency asymptotically. However, for a given real dataset, the performance of $\widehat{\theta}_{\lambda_T}$ can be sensitive to $\lambda_T$ selected from this range, as seen in Fig.~\ref{fig::toy example composition loss} earlier. Hence, using this range only to select $\lambda_T$ is not practical. To this end, we provide a data-driven approach to select the tuning $\lambda_T$,  and we theoretically show that using the data-driven approach to select $\lambda_T$ can also leads to selection consistency. It is seen in \S\ref{sect::simulation}, \ref{sect::real data} and \S\ref{sub sect::simulation network}, \ref{sub sect::simulation function} in the Supplementary Material that it provides reliable performance of period estimation for different finite settings in numerical studies.

To establish the theory of the information criterion, we partition $\lambda$
into three mutually exclusive sets $\Lambda_0=\{\lambda:\widehat{\theta}_\lambda=\theta_0\}$, $\Lambda_+=\{\lambda:\widehat{\theta}_\lambda\in\mathcal{M}_{\theta_0},\widehat{\theta}_\lambda\neq\theta_0\}$ and $\Lambda_-=\{\lambda:\widehat{\theta}_\lambda\notin\mathcal{M}_{\theta_0}\}$, where  $\mathcal{M}_{\theta_0}=\{\theta: \theta = k\theta_0,k=1,2,\ldots,1\leq \theta \leq \Theta_T\}$. One can see that the sets $\Lambda_0$, $\Lambda_+$ and $\Lambda_-$ correspond to whether the resulting model $\widehat{m}_{t}(\cdot,T)$ is correctly fitted, overfitted or misspecified. 

\begin{theorem}\label{thm::IC}
        Assume that $\Theta_T\to\infty$ and $(\Theta_T\log \Theta_T)/T\to 0$ as $T\to\infty$.  
        Then under Assumptions~\ref{assumption 1}--\ref{assumption 5}, \begin{align}\label{eq::thm IC statement}
            {\rm pr}\left( \min_{\lambda\in\Lambda_+ \cup \Lambda_-} {\rm IC}_\lambda > {\rm IC}_{\lambda_T} \right) \to 1
        \end{align}
        holds as $T\to\infty$ if (i) $g(T)\gg \{T/(\Theta_T\log \Theta_T)\}^{-1/(\beta^\prime-1)}$ and (ii) $g(T)=o(1)$ as $T\to\infty$ for any reference tuning parameter $\lambda_T$ satisfying $T^{1-1/(\beta^\prime-1)}(\Theta_T\log\Theta_T)^{1/(\beta^\prime-1)}\ll \lambda_T \ll T$, for any $\beta^\prime>\beta$, with $\beta$ given in  Assumption~\ref{assumption 4}. 
\end{theorem}
\begin{remark}
     When $\Theta_T\to\infty$, the cardinality of $\lambda\in\Lambda_+ \cup \Lambda_-$ is not finite and the pointwise result ${\rm pr}\left( {\rm IC}_\lambda > {\rm IC}_{\lambda_T} \right) \to 1$ for every $\lambda\in\Lambda_+ \cup \Lambda_-$ does not imply (\ref{eq::thm IC statement}). Thus we need the uniform result for $\widehat{m}_t(\theta,T)$ over $\theta\in\{1,2,\ldots,\Theta_T\}$.
\end{remark}

Theorem \ref{thm::IC} indicates that any $\lambda$ failing to identify the true period cannot be selected as the optimal tuning parameter. To see this, note that by Theorem \ref{thm::period estimation consistency}, we have $\widehat{\theta}_{\lambda_T}=\theta_0+o_p(1)$ as $T\to\infty$ when $\lambda_T$ satisfying the condition in Theorem \ref{thm::period estimation consistency}. Therefore, Theorem \ref{thm::IC} indicates that if $\lambda$ falls into the set $\Lambda_+$ or $\Lambda_-$, then the corresponding ${\rm IC}_\lambda$ will be larger than the reference level ${\rm IC}_{\lambda_T}$ with probability tending to 1 as $T\to\infty$. For $\widehat{\lambda}=\argmin_\lambda {\rm IC}_\lambda$, we always have ${\rm IC}_{\widehat{\lambda}}\leq {\rm IC}_{\lambda_T}$, which can ensure that $\widehat{\lambda}$ falls into the set $\Lambda_0=\{\lambda:\widehat{\theta}_\lambda=\theta_0\}$. Moreover, as we allow $\Theta_T\to\infty$ as $T\to\infty$, this means the range of the candidate models diverges and it is not surprising that the regularization function $g(T)$ involves $\Theta_T$, see, e.g., \cite{chen2008extended} and \cite{zhang2025penalty}. Notably, when $\Theta_T=\Theta$ is fixed and for metric spaces that satisfy $\beta=2$ given in Assumption~\ref{assumption 4}, see Remark \ref{remark::1}, Theorem \ref{thm::IC} leads to a specific consideration of the information criterion which is the classical Bayesian information criterion (\citealt{schwarz1978estimating}) as ${\rm BIC}_\lambda = \log \{ {\rm RSS}(\widehat{\theta}_\lambda) / T \} + \widehat{\theta}_\lambda \log(T) /T$.

Combining Theorem \ref{thm::period estimation consistency} and Theorem \ref{thm::IC} establishes the consistency of $\widehat{\theta}_{\widehat{\lambda}}$ with $\widehat{\lambda}$ selected via (\ref{formula::lambda selection}),  as stated in Corollary \ref{cor::final period est} below.

\begin{corollary}\label{cor::final period est}
    Suppose the conditions in Theorem \ref{thm::IC} hold.  Then $\widehat{\theta}_{\widehat{\lambda}}-\theta_0=o_p(1)$ with $\widehat{\lambda}$ given by (\ref{formula::lambda selection}).
\end{corollary}

\subsection{Theoretical Result for the Periodic Component Estimator}

We now establish the convergence rate of the periodic component estimator $\widehat{m}(t)$ for $t=1,\ldots,T$ given by (\ref{formula::periodic component estimator}).

\begin{theorem}\label{thm::periodic component est consistency}
    Under Assumptions~\ref{assumption 1}--\ref{assumption 5},
    \[
    \max_{t=1,\ldots,T} d(\widehat{m}(t),m(t)) = O_p\left( T^{-1/2(\beta^\prime-1)} \right),
    \]
    for any $\beta^\prime>\beta$, with $\beta$ given in  Assumption~\ref{assumption 4}. 
\end{theorem}

Theorem~\ref{thm::periodic component est consistency} presents a uniform convergence result. The faster uniform convergence rate, compared to the one in Proposition~\ref{prop::uniform convergence rate partial mean}, is achieved because  we only need the pointwise convergence rate of $d(\widehat{m}(t),m(t))$. To see this, note that the maximum over $1,2,\ldots,T$ is the same as over $1,2,\ldots,\widehat{\theta}_{\widehat{\lambda}}$ while we have $\widehat{\theta}_{\widehat{\lambda}}=\theta_0$ with probability tending to 1 as $T\to\infty$ by Theorems~\ref{thm::period estimation consistency} and \ref{thm::IC}.

\section{Simulation Studies}\label{sect::simulation}

In this section, we illustrate our method by the simulation of periodic spherical data.  Additional simulation results for networks and distributional data can be found in \S\ref{sub sect::simulation network} and \ref{sub sect::simulation function} in the Supplementary Material. For each scenario, we replicate the simulation 200 times with sample size $T$ varying in $\{100,240,500\}$ and with three different levels of noise. The tuning parameter $\lambda_T$ is chosen by minimizing the information criterion in (\ref{eq::BIC}), with the regularization function $g(T)$ taking the form of $\{(\Theta_T\log \Theta_T)/T\}^{-0.5}$, where $\Theta_T$ is allowed to grow with the sample size $T$. Note that this choice of  $g(T)$ satisfies the conditions in Theorem~\ref{thm::IC}.

\subsection{Periodic Spherical Data Generation}\label{supp sect:: composition generate}

The generation of periodic spherical time series is constructed based on the spherical autoregressive model in \cite{zhu2024spherical}. Choose the metric $d$ as the great circle distance, for $\nu_1,\nu_2\in\mathcal{S}$, define the spherical difference based on the skew-symmetric operator as $\nu_1\ominus\nu_2= \eta (\zeta_2\otimes \zeta_1 - \zeta_1\otimes\zeta_2)$ where $\eta=d(\nu_2,\nu_3)$, $\zeta_1=\nu_2$, $\zeta_2=(\nu_1-\langle\nu_1,\nu_2\rangle\nu_2)/\|\nu_1-\langle\nu_1,\nu_2\rangle\nu_2\|_{\rm E}$ with $\langle\cdot,\cdot\rangle$ be the inner product and $\|\cdot\|_{\rm E}$ be the Euclidean norm, and $\otimes$ denoting the tensor product. Extending the spherical autoregressive model in \cite{zhu2024spherical} to the periodic spherical autoregressive model as 
\begin{align}\label{model::spheircal AR model}
    \varpi_t - \mu_{\varpi} = \varphi_1 (\varpi_{t-1} - \mu_{\varpi}) + \cdots + \varphi_p (\varpi_{t-p} - \mu_{\varpi}) + \epsilon_t
\end{align}
where $\Xi_t=Y_t\ominus m(t)$, $\mu_{\varpi}=\mathbb{E}(\varpi_t)$, $\varphi_1,\varphi_2,\ldots,\varphi_p\in\mathbb{R}$ are the spherical autoregressive coefficients, and $\{\epsilon_t\}_{t=1}^T$ are independent and identically distributed random noise with mean 0.

For a given sample size $T$, an autoregressive order $p$, and the autoregressive coefficients $\{\varphi_j\}_{j=1}^p$, we generate the skew-symmetric operator data following the period autoregressive structure in (\ref{model::spheircal AR model}). The data generation mainly follows \cite{xu2026spherically} where we first generate $p$ skew-symmetric operators $\{\widetilde{\varpi}_k=\widetilde{Y}_k\ominus m(k)\}_{k=1}^p$ where $\widetilde{Y}_k^{(2)}\in\mathcal{S}^6$  and are generated independently and identically distributed by the von Mises--Fisher distribution with the mean direction being $m(t)$ and the concentration parameter being $7\alpha$ where $\alpha\in\{2,3,4\}$ controls the noise level. The noise level decreases as $\alpha$ grows. Then, for each $t=p+1,p+2,\ldots,(T+p+500)$, we generate $\widetilde{\varpi}_t = \sum_{j=1}^p \varphi_j \widetilde{\varpi}_{t-j} + \widetilde{\epsilon}_t$ by considering $\mu_{\varpi}=0$ and $\widetilde{\epsilon}_t=e_t\ominus \mu_{R,2}$ where $e_t\in\mathcal{S}^{m-1}$ is independently and identically distributed generated by the von Mises--Fisher distribution with the mean direction being $\mu_c=I_7/\|I_7\|_{\rm E}$ and the concentration parameter being $7\alpha$. Here, $I_7$ denotes a vector of the ones with dimension 7. We then pick $\varpi_t=\widetilde{\varpi}_{t+p+500}$ for every $t=1,2,\ldots,T$. The periodic component $m(t)$ is constructed as follows. We first identify an orthonormal basis for the orthogonal complement of $\mu_c$, denoted as $\{\nu_1, \nu_2\} \in {\rm null}(\mu_c^\perp)$. For $t=1,2,\dots,\theta_0$, $m(t)$ is generated as $m(t) = \cos(\psi)\mu_c + \sin(\psi)\left[ \cos(\phi_t)\nu_1 + \sin(\phi_t)\nu_2 \right]$ where $\psi=0.3$ represents the angular radius of the component from the mean $\mu_c$, $\phi_t = 2\pi t/\theta_0$ is the phase angle at time $t$, ensuring that the periodic component $m(t)$ completes exactly one revolution over the period $\theta_0$. We also set $m(l)=m(l+\theta_0)=m(l+2\theta_0)=\cdots$ for $l\in\{1,\ldots,\theta_0\}$. In the numerical studies, we fix the true period $\theta_0=12$, the autoregressive order $p=1$, and the autoregressive coefficient $\varphi_1=0.5$. The final periodic spherical time series $\{Y_t\}_{t=1}^T$ is generated via $Y_t={\rm Exp}(\varpi_t)m(t)$ where ${\rm Exp}$ denotes the exponential map.

Figure~\ref{fig::periodic spherical ts} in the Supplementary Material visualizes the generated periodic spherical time series for $\alpha\in\{2,3,4\}$. While it is hard to directly observe a clear periodic pattern directly based on Fig.~\ref{fig::periodic spherical ts}, the numerical results below show that the proposed method produce consistent estimation of the true but unknown period.

\subsection{Simulation Results}\label{sub sect::simulation composition}

In this simulation, the range of the period candidates is set to grow at the rate of $T^{1/3}$. When $T=100$, $\Theta_T=37$ which covers three full periods, since the true period is 12 and $3 \times 12 < \Theta_T=40$. When $T=240$, $\Theta_T=49$, while when $T=500$, $\Theta_T$ is set to be 63. Denote $p(\widehat{\theta}_{\widehat{\lambda}}=a)$ as the proportion of cases where $\widehat{\theta}_{\widehat{\lambda}}=a$ across 200 replicates, and  $p(b\leq \widehat{\theta}_{\widehat{\lambda}}\leq c)$ as the proportion of cases where $b\leq \widehat{\theta}_{\widehat{\lambda}}\leq c$ across 200 replicates, for some integers $a,b$ and $c$.

\begin{table}[!htbp]
\caption{Empirical probabilities that $\widehat{\theta}_{\widehat{\lambda}}=12$, reported in the first three columns, and that $8 \leq \widehat{\theta}_{\widehat{\lambda}}\leq 16$, reported in the last three columns, for the sample size $T$ varies in $\{100,240,500\}$ and different levels of Fr\'{e}chet variance of each observation when the simulated samples are periodic spherical time series and the information criterion is for the logarithm of residual sum of squares. \label{table::composition simulation period est criteria 1}}
\centering
\scalebox{0.9}{
\begin{tabular}{l | ccc | ccc }
  \hline
&\multicolumn{3}{c|}{$p(\widehat{\theta}_{\widehat{\lambda}}=12)$}  & \multicolumn{3}{c}{$p(8 \leq \widehat{\theta}_{\widehat{\lambda}}\leq 16)$}  \\
 & $T=100$ & $T=240$ & $T=500$ & $T=100$ & $T=240$ & $T=500$ \\ 
  \hline
$\alpha = 2$ & 0.655 & 0.975 & 1.000 & 0.690 & 0.975 & 1.000 \\ 
  $\alpha = 3$ & 0.820 & 1.000 & 1.000 & 0.820 & 1.000 & 1.000 \\ 
  $\alpha = 4$ &  0.870 & 1.000 & 1.000 & 0.870 & 1.000 & 1.000 \\
    \hline
\end{tabular}
}
\end{table}

Table \ref{table::composition simulation period est criteria 1} presents the simulation results for periodic compositions using the first IC in (\ref{eq::BIC}). The first three columns of the table report the probabilities that the estimator $\widehat{\theta}_{\widehat{\lambda}}$ correctly identifies the true period $\theta_0=12$, while the last three columns of the table show the probabilities of $\widehat{\theta}_{\widehat{\lambda}}$ ranging from 8 to 16. Notably, our proposed estimator exhibits high accuracy even with a small sample size when the Fr\'{e}chet variance of $Y_t$ is small. When the variance becomes large, it is not surprising that the finite sample performance of the estimator becomes worse for small sample sizes. Nonetheless, the estimator typically provides a good approximation of the true period in most cases. Its accuracy increases rapidly with the number of observed cycles of the periodic pattern.

\begin{table}[!htbp]
\caption{Empirical probabilities that $\widehat{\theta}_{\widehat{\lambda}}=12$, reported in the first three columns, and that $8 \leq \widehat{\theta}_{\widehat{\lambda}}\leq 16$, reported in the last three columns, for the sample size $T$ varies in $\{100,240,500\}$ and different levels of Fr\'{e}chet variance of each observation when the simulated samples are periodic spherical time series and the information criterion is for the residual sum of squares. \label{table::composition simulation period est criteria 2}}
\centering
\scalebox{0.9}{
\begin{tabular}{l | ccc | ccc }
  \hline
&\multicolumn{3}{c|}{$p(\widehat{\theta}_{\widehat{\lambda}}=12)$}  & \multicolumn{3}{c}{$p(8 \leq \widehat{\theta}_{\widehat{\lambda}}\leq 16)$}  \\
 & $T=100$ & $T=240$ & $T=500$ & $T=100$ & $T=240$ & $T=500$ \\ 
  \hline
$\alpha = 2$ & 0.655 & 0.975 & 1.000 & 0.690 & 0.975 & 1.000 \\ 
  $\alpha = 3$ & 0.820 & 1.000 & 1.000 & 0.820 & 1.000 & 1.000 \\ 
  $\alpha = 4$ &  0.870 & 1.000 & 1.000 & 0.870 & 1.000 & 1.000 \\
    \hline
\end{tabular}
}
\end{table}

Table~\ref{table::composition simulation period est criteria 2} provides the simulation results for periodic compositions using the second IC in (\ref{eq::BIC}). The results exhibit the same patterns to those in Table \ref{table::composition simulation period est criteria 1}. When the sample size $T=500$, the estimator always hits the value of the true period even when the variance of each element of $Y_t$ is larger such as $\alpha=10$.

\begin{figure}[!htbp]
\centering
\includegraphics[width=1\linewidth]{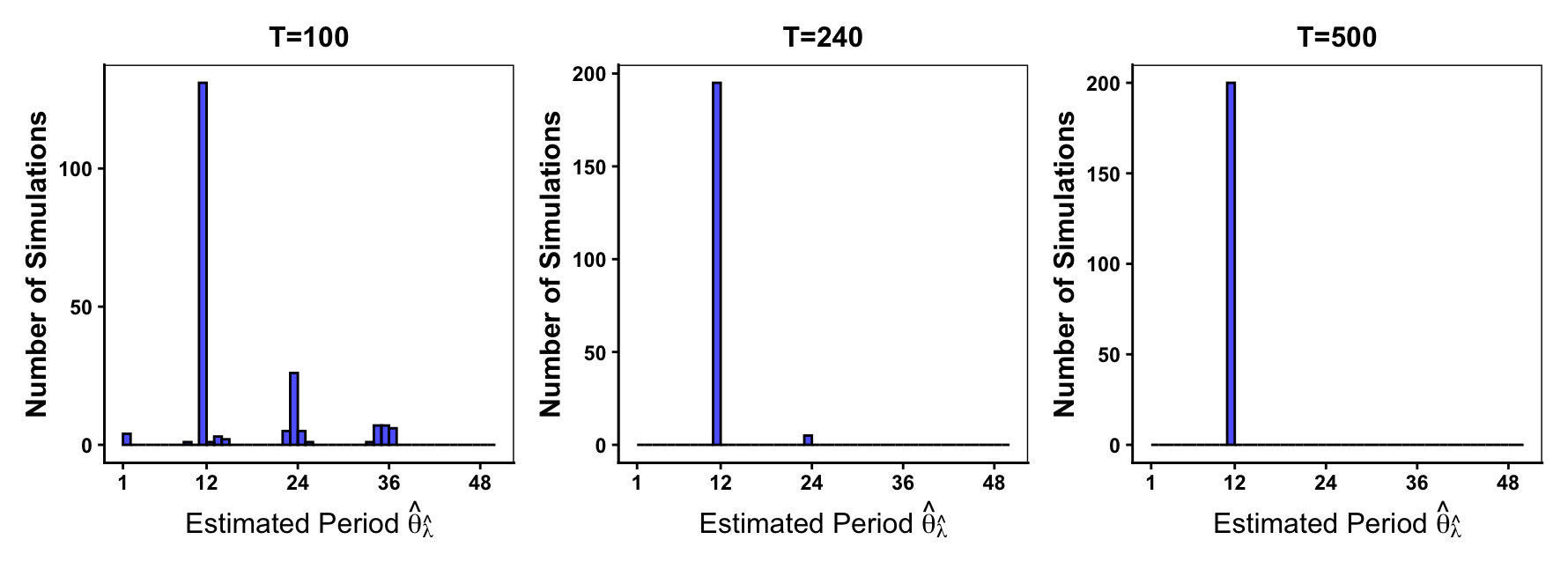}

\caption{ Histograms of the simulation results for periodic compositions when $\alpha = 2$ for different samples sizes $T\in\{100,240,500\}$. The bars present how many times each value of $\widehat{\theta}_{\widehat{\lambda}}$ is observed across 200 simulation runs using the information criterion for the residual sum of squares.  \label{fig::composition simulation period est hist}}
\end{figure}

Histograms of the simulation results when $\alpha = 2$ using the second IC in (\ref{eq::BIC}) for different sample sizes $T\in\{100,240,500\}$ are given in Fig.~\ref{fig::composition simulation period est hist}. Each panel presents the distribution of $\widehat{\theta}_{\widehat{\lambda}}$ for $T=100,240,500$. The figure reveals that for a small sample size $T=100$, alongside a primary concentration of estimates around the true period $\theta_0$, secondary clusters emerge around multiples of $\theta_0$. Although Table \ref{table::composition simulation period est criteria 2} suggests that the estimator may not always precisely recover the true period when variances are large, it still provides reasonable approximations, particularly capturing its multiples. As suggested by the proof of Theorem \ref{thm::period estimation consistency}, this pattern aligns with the asymptotic behaviour of $\widehat{\theta}_{\widehat{\lambda}}$. When the sample size increases, the clusters around the multiples of $\theta_0$ disappear, leading to more accurate period identification. When the sample size $T=500$, the proposed estimator always correctly identifies the value of the true period.

\begin{figure}[!htbp]
\centering
\includegraphics[width=1\linewidth]{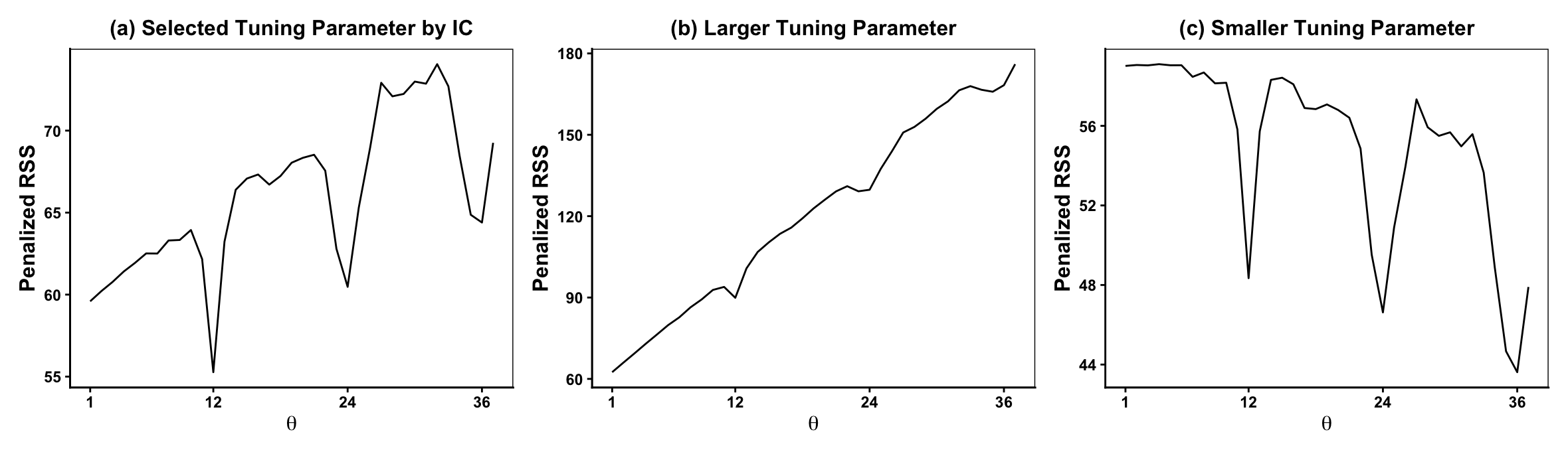}

\caption{ Plots of penalized RSS for one simulation replicate  for periodic compositions with $T=100$ and $\alpha = 4$, considering different choices of the tuning parameter $\lambda_T$ such as $\lambda_T=\widehat{\lambda}$ in (a), $\lambda_T=5\widehat{\lambda}$ in (b), and $\lambda_T=\widehat{\lambda}/5$ in (c), using the information criterion for the residual sum of squares. \label{fig::simulation composition loss}}
\end{figure}

Now we study the performance of our selected tuning parameter based on the proposed IC in (\ref{eq::BIC}). We conduct a typical simulation for periodic compositions with $T=100$ and $\alpha = 1$. Figure~\ref{fig::simulation composition loss} presents the penalized RSS $\mathcal{L}(\theta,\lambda_T)$ from (\ref{formula::loss function}) for three selected values of tuning parameter $\lambda_T$ where $\lambda_T=\widehat{\lambda}$ in panel (a), $\lambda_T=5\widehat{\lambda}$ in panel (b) and $\lambda_T=\widehat{\lambda}/5$ in panel (c). Panels (a) and (c)  in Fig.~\ref{fig::simulation composition loss} show that sharp decreases in the penalized RSS are observed near the true value of the period $\theta_0=12$ and its multiples, when the tuning parameter is not too large. Moreover, the choice of $\lambda_T$ significantly affects the function’s overall trend where a larger tuning parameter forces the global minimum to be 1 in panel (b) while a smaller tuning parameter causes that the global minimum occurs at a later spike instead of at the true period. These findings support the conclusion that our selected tuning parameter based on the proposed IC performs in a reasonable manner.

The final step of our proposed methodology is the estimation of the periodic component $m(t)$ of the random object, given the period estimation. We denote by $\widehat{m}^{(k)}(t)$  the estimated component for the $k$-th replicate and we calculate the mean of squared distance (MSE) for the $k$-th replicate between $\widehat{m}^{(k)}(t)$ and $m(t)$, denoted as ${\rm MSE}_k=T^{-1}\sum_{t=1}^T d^2(\widehat{m}^{(k)}(t),m(t))$, where the distance is measured by the geodesic distance after the square root transformation. The final MSE is computed as $200^{-1}\sum_{k=1}^{200} {\rm MSE}_k$.

\begin{table}[!htbp]
\caption{Mean squared errors (MSEs) for periodic component estimation considering different sample sizes $T\in\{100,240,500\}$ and different levels of Fr\'{e}chet variance of each observation when the simulated samples are periodic compositions. \label{table::composition simulation period component est}}
\centering
\scalebox{1}{
\begin{tabular}{l | ccc  }
  \hline
 & $\alpha = 2$ &  $\alpha = 3$ &  $\alpha = 4$\\ 
  \hline
$T=100$ & 0.088 & 0.053 & 0.039 \\ 
$T=240$ & 0.039 & 0.023 & 0.016 \\ 
$T=500$& 0.019 & 0.011 & 0.008 \\ 
  \hline
\end{tabular}
}
\end{table}

Table \ref{table::composition simulation period component est} shows the MSE of the periodic component estimation considering different sample sizes $T\in\{100,240,500\}$ and different choices of $\alpha$, corresponding to the different levels of Fr\'{e}chet variance of each observation. For each choice of $\alpha$, one can see that when the sample size increases, the MSE becomes smaller. Notably, for small sample size $T=100$, the period estimation does not hit the true period all the time, resulting in a larger MSE. For larger sample sizes $T\in\{240,500\}$, the period estimation accurately reach the true period, and thus gives smaller MSE. When $\alpha$ becomes smaller, the Fr\'{e}chet variance of each observation gets larger and one can observe that the MSE becomes larger for the same sample size.

Table \ref{table::composition simulation period component est}  supports the finding that the finite sample performance of the periodic component estimator $\widehat{m}(t)$ for $t=1,\ldots,T$ depends on the finite performance of the period estimator $\widehat{\theta}_{\widehat{\lambda}}$. While the period estimator always hits the true period value, Table \ref{table::composition simulation period component est} aligns with the numerical results in \cite{petersen2019frechet} and \cite{zhou2022network}. Therefore, given a good estimation of $\theta_0$, the estimator $\widehat{m}(t)$ can be expected to perform similarly to the standard estimator of the global Fr\'{e}chet regression for $t=1,2,\ldots,T$. For this reason, we focus on the properties of $\widehat{\theta}_{\widehat{\lambda}}$ in the remaining simulation studies for networks and distributions in the Supplementary Material.

\section{Real Data Analysis}\label{sect::real data}

\subsection{U.S. Electricity Generation }

We analyse the monthly composition of U.S. electricity generation using data from \url{https://www.eia.gov/electricity/data/browser/}. The data is preprocessed into a compositional form, where each entry in the compositional vector represents the percentage contribution of a specific energy source to net electricity generation. Following the same preprocessing idea in \cite{dubey2023change}, we consolidate similar resource categories, resulting in seven broad groups: Coal; Petroleum (including petroleum liquids and petroleum coke); Gas (natural and other gases); Nuclear; Conventional hydroelectric; Renewables (wind, geothermal, biomass, and other renewables); and Solar (encompassing small-scale solar photovoltaic and all utility-scale solar). Our dataset consists of $T=180$ monthly observations spanning from January 2001 to December 2015. Notably, each observation $Y_t$ takes values a six-simplex $\Delta^6=\{\bm{\delta}=(\delta_1,\ldots,\delta_7)^\top\in\mathbb{R}^7:\sum_{l=1}^7 \delta_l = 1~{\rm and}~\delta_p\geq0~{\rm for}~p=1,\ldots,7\}$. Here we consider the square root transformation for compositional data (\citealt{scealy2011regression}), resulting in spherical data, and choose the corresponding metric be the geodesic distance; that is, we use the metric given by $d(\bm{s}_1,\bm{s}_2) = \arccos(\bm{s}_1^\top\bm{s}_2)$ where $\bm{s}_i=\sqrt{\bm{\delta}_i}$ for $i=1,2$ with $\bm{\delta}_i$ being a compositional data vector and $\sqrt{\cdot}$ being the positive component-wise square root.

\begin{figure}[!htbp]
\centering
\includegraphics[width=1\linewidth]{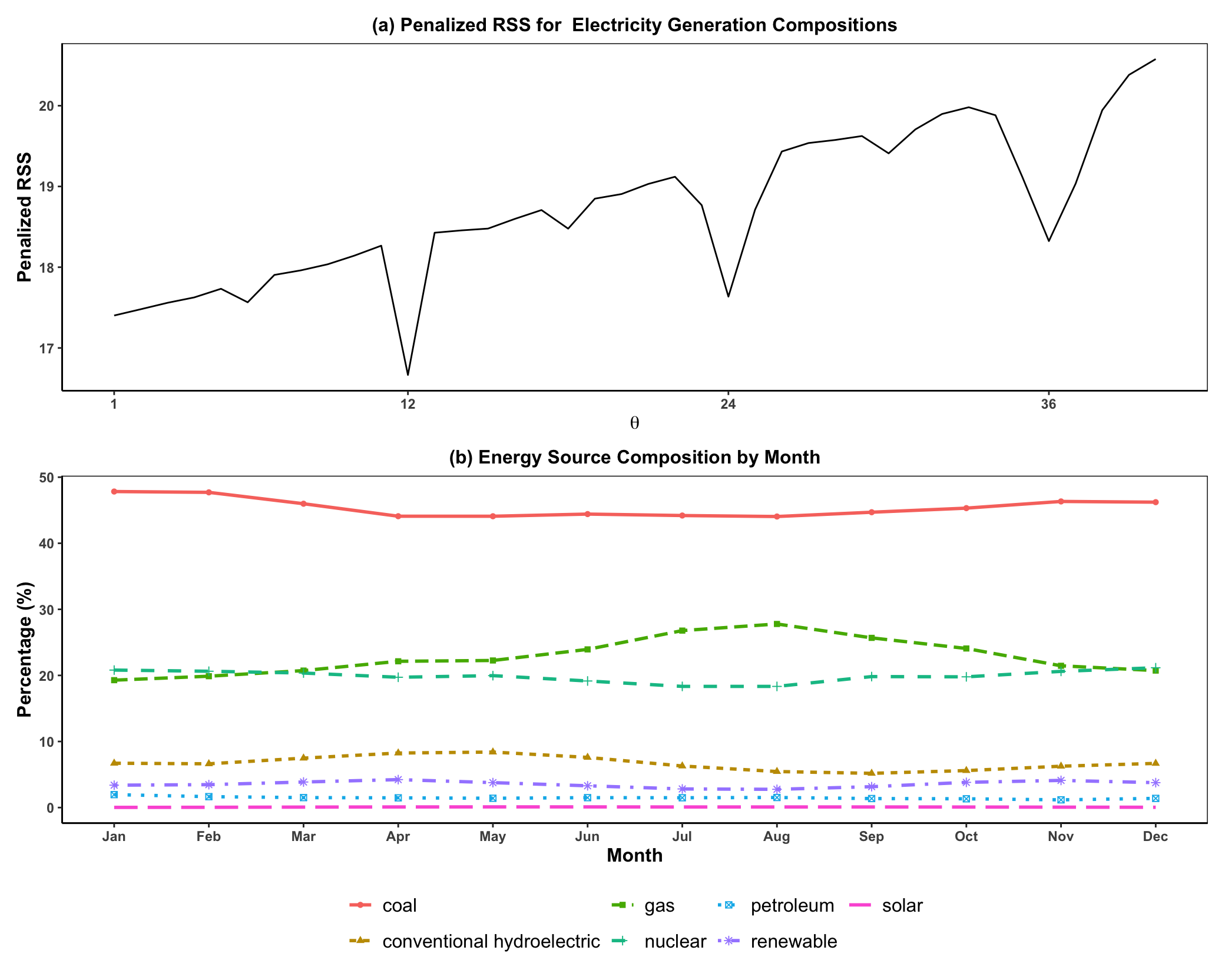}

\caption{(a) Plot of the penalized RSS with the selected tuning parameter for the monthly electricity generation compositions in U.S. (b) Plot of the elements of the estimated periodic component of the monthly electricity generation compositions in U.S..  \label{fig::real data period composition}}
\end{figure}

Applying our proposed methodology, we identify a periodicity of 12 months in U.S. monthly electricity generation compositions, as shown in Fig.~\ref{fig::real data period composition}(a). Figure~\ref{fig::real data period composition}(a) shows that the penalized RSS exhibits distinct sharp drops around the period 12 and its multiples, indicating a yearly periodic pattern in electricity generation compositions. Although total energy generation is known to exhibit an annual periodic pattern, peaking in summer and winter due to elevated demand, the composition of energy sources does not necessarily share this yearly cycle. Indeed, it is entirely possible for the relative proportions of each energy source to remain constant, even as the aggregate generation volume fluctuates periodically.

To investigate the reason why the generation compositions have a yearly cycle, we next turn to the estimation of the periodic component $m(t)$ for $t=1,\ldots,\theta_0$. Using the estimated period $\widehat{\theta}_{\widehat{\lambda}}=12$, we obtain the estimation via (\ref{formula::periodic component estimator}). Figure~\ref{fig::real data period composition}(b) illustrates the evolution of each energy source's percentage based on the estimation over time. The figure reveals clear seasonal trends in energy composition and provides meaningful insights. Notably, the percentage of gas peaks between July and September, aligning with increased summer demand and the fact that natural gas has become the dominant source of electricity generation in the U.S. Meanwhile, coal consistently reaches its highest share at the beginning of each year. Conventional hydroelectric and renewable energy sources show higher percentages around April and lower around August. Moreover, the proportion of nuclear decreases in July and August, mainly due to the higher temperature in summer and nuclear power plants relying on water for cooling. In the summer, higher water temperatures can reduce cooling efficiency, forcing plants to lower output to maintain safe operating conditions.

\subsection{New York City Citi Bike Sharing System}

The New York City Citi Bike sharing system contains historical bike trip data, accessible at \url{https://citibikenyc.com/system-data}. This public dataset records trip start and end times, along with corresponding locations, at a second-resolution level, covering rides between bike stations throughout New York City.

Our study focuses on trips recorded over 15 weekdays in November 2019 to analyse the periodicity of time-varying transportation networks. By examining the hourly dynamics of bike rides across various stations, we aim to identify periodic patterns within the Citi Bike system and broader transportation trends. We consider the 90 most frequently used bike stations and segment each day into 24 one-hour intervals. For each one-hour interval, we construct a network with 90 nodes representing the selected stations, where edge weights indicate the number of trips between station pairs. This approach results in a time-varying network spanning 15 weekdays, from November 4 to November 22, 2019, yielding a total of 360 observations.

Each observation corresponds to a $90 \times 90$ graph Laplacian matrix that encapsulates the network among the 90 bike stations for a given one-hour period. For a network with 90 nodes, the graph Laplacian matrix $L$ is obtained as $L = D - A$, where $A$ is the $90 \times 90$ adjacency matrix, with the $(i,j)$-th element $a_{ij}$ being the edge strength between stations $i$ and $j$, and $D$ is a diagonal matrix with diagonal elements $d_{ii} = \sum_{j=1}^p a_{ij}$. The graph Laplacian provides a distinctive representation of the network structure. The metric we consider for the network Laplacians is the Frobenius metric.

\begin{figure}[!htbp]
\centering
\includegraphics[width=1\linewidth]{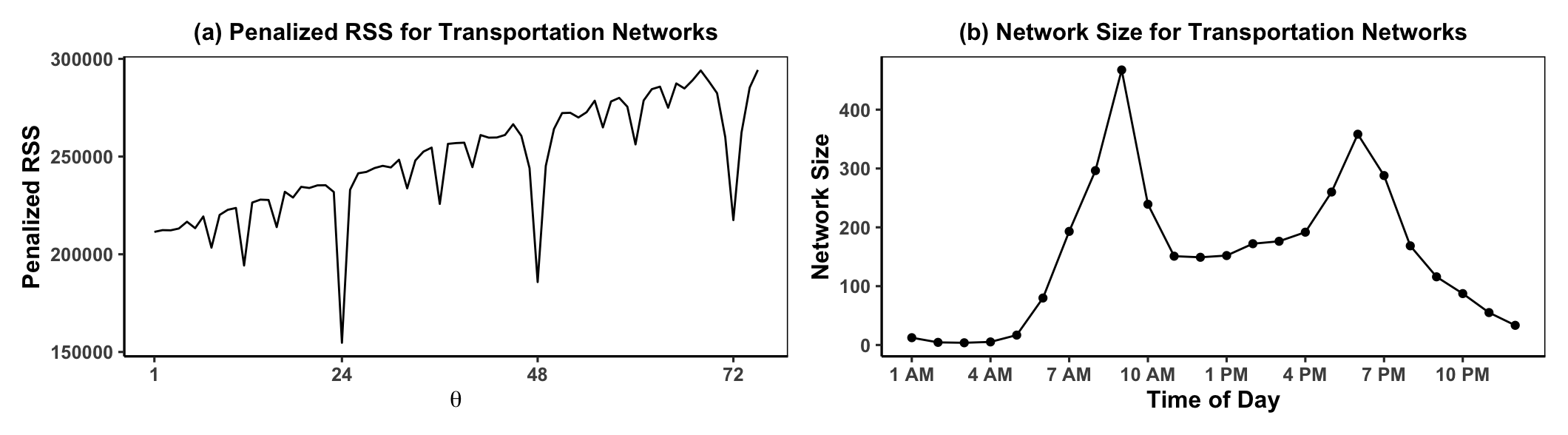}

\caption{(a) Plot of the penalized RSS with the selected tuning parameter for hourly transportation networks in New York City Citi Bike Sharing system. (b) Plot of the network sizes of the estimated periodic component of the hourly transportation networks in New York City Citi Bike Sharing system.  \label{fig::real data period network}}
\end{figure}

In what follows, we apply our proposed methodology to estimate the unknown period of the transportation networks. Once the period is estimated, we proceed to estimate the periodic component of the transportation networks. Figure~\ref{fig::real data period network}(a) displays the penalized RSS with the selected tuning parameter for the hourly transportation data. The most notable feature is the noticeable dips with a global minimum at $\theta=24$, along with additional spikes at its multiples. As discussed in \S\ref{subsection::period estimation} and \ref{sect::simulation}, such spikes are indicative of the presence of a periodic pattern in the data. The sharpness of the spike and the shape of the penalized RSS strongly suggest a periodic pattern with a period of 24 in the hourly transportation networks of the New York City Citi Bike Sharing system. Figure~\ref{fig::real data period network}(a) indicates that there exists a daily pattern for the hourly transportation networks, which has been observed in \cite{xu2025change}.

To further analyse the periodic component, we consider to visualize the estimated periodic component by calculating the network size of the estimated periodic component at each time $t$, as illustrated in Fig.~\ref{fig::real data period network}(b). The figure reveals distinct peaks at 9 am and 6 pm, corresponding to morning and evening commuting surges, respectively. This result demonstrates that our proposed methodology effectively captures and quantifies periodicity in hourly transportation networks, as further supported by the interpretable patterns in Fig.~ \ref{fig::real data period network}(b).

\subsection{Water Consumption in Germany}\label{subsect::water application}

We analyse daily functional water consumption in Germany using a dataset containing time-resolved measurements of water usage (in cubic meters) recorded hourly from January 2016 to March 2016. These measurements were collected at a pumping station within a regional water supply network, serving a mix of residential, commercial, and industrial consumers. To facilitate analysis, we preprocess the daily observations into a functional form, where each function represents the hourly water consumption pattern for a given day. After the preprocessment, the data consists of $T=70$ daily water consumption curves. Additionally, we focus on the shape of daily water consumption curves and normalize every curves, resulting in distributional time series. The metric we use after the normalization is the Wasserstein metric, same with the one used in \S\ref{sub sect::simulation function} in the Supplementary Material.

\begin{figure}[!htbp]
\centering
\includegraphics[width=1\linewidth]{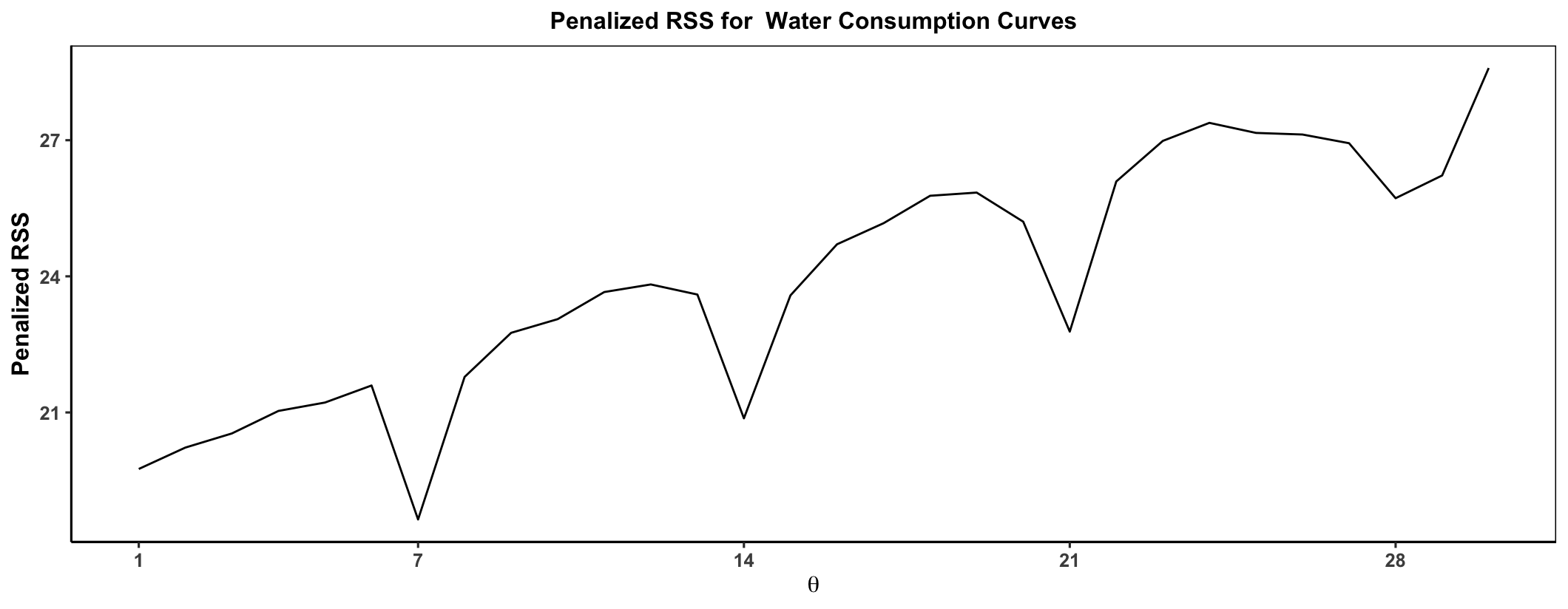}

\caption{Plot of the penalized RSS with the selected tuning parameter for the daily water consumption curves in Germany.  \label{fig::real data period function loss}}
\end{figure}

After employing the proposed methodology, a period of 7 has been identified based on the penalized RSS shown in Fig.~\ref{fig::real data period function loss}. In this figure, we observe sharp decreases around the period 7 and its multiple, indicating a weakly periodic pattern in water consumption curves.

\begin{figure}[!htbp]
\centering
\includegraphics[width=1\linewidth]{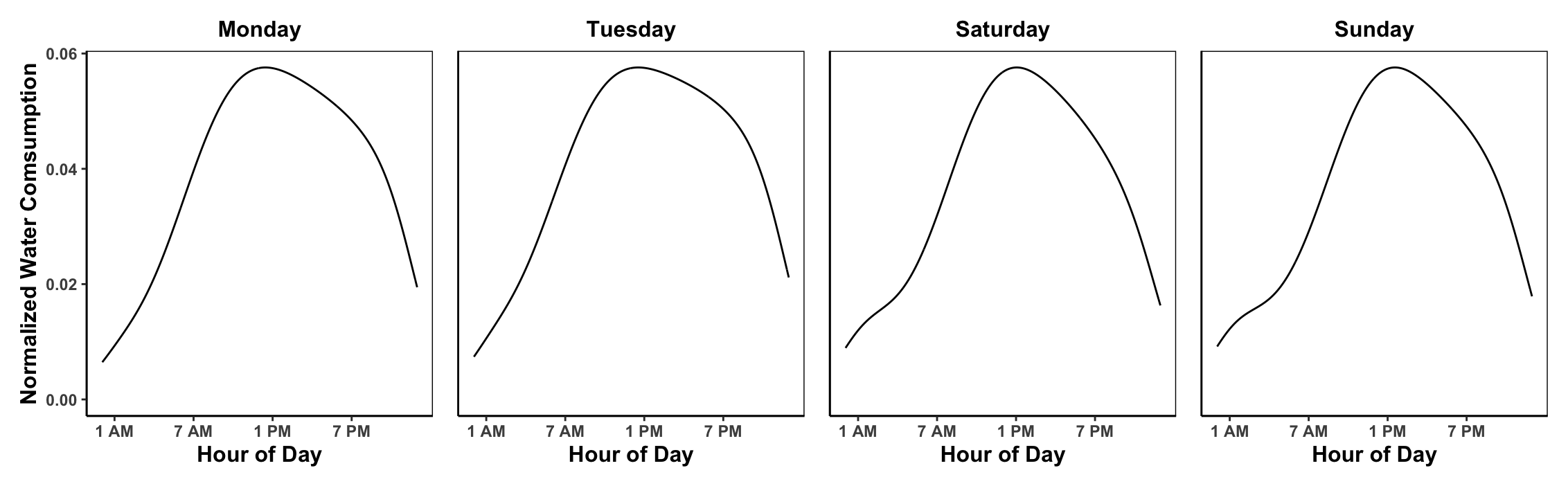}

\caption{Plots of the part of the estimated periodic component of the normalized daily water consumption curves in Germany. The estimated periodic component has a period 7 and contains 7 curves while 4 curves are given here. \label{fig::real data period function component est sub}}
\end{figure}

Turning to the periodic component estimation, Fig.~\ref{fig::real data period function component est sub} presents 4 of 7 curves in the estimated periodic component, the first two represent the weekday water consumption curves while the last two represent the weekend curves. The plots of all the elements in the estimated periodic component can be found in Fig.~\ref{fig::real data period function component est} in the Supplementary Material. A clear periodic pattern can be observed in Fig.~\ref{fig::real data period function component est sub}. Notably, the daily water consumption curves have similar shapes on weekdays and weekends, respectively. For weekday consumption curves, peaks occur before 1 pm while the peaks of weekend consumption curves happen after 1 pm. Notably, water consumption exhibits a smooth increasing trend starting early in the morning on weekdays while the  increase appears less steep for weekends in the morning, especially on Sunday. The variations in morning and early afternoon water consumption are consistent with the differences in daily routines between workdays and weekends.

\section{Discussion}

A fundamental open question in the statistical analysis of random objects is the optimal choice of metric. For instance, when analyzing spherical data on the unit sphere, one must choose between an intrinsic metric, such as the great circle distance, and an extrinsic one, such as the induced Euclidean distance. As demonstrated in \eqref{formula::periodic component}, our proposed periodicity quantification framework relies heavily on this choice. Consequently, establishing a data-driven approach for metric selection remains a compelling direction for future research. Furthermore, if the object-valued time series $Y_1, Y_2, \ldots, Y_T$ is embedded into a Euclidean space $\mathbb{R}^p$, one might naturally consider an extrinsic analysis by extending the methodology of \cite{vogt2014nonparametric}. However, as illustrated in \S~\ref{supp sect::failure} of the supplementary material, such extrinsic approaches can fail to recover the true underlying periodic pattern if the induced Euclidean metric is unsuited for the data. This also underscores the critical role of metric selection in random object analysis.

\section*{Supplementary material}
\label{SM}
The supplementary material consists of limitations of extrinsic methods; proofs of theorems propositions and corollaries;  an additional figure for the simulation on spherical data; additional simulation studies on network Laplacians and distributions; and an additional figure for the case study.

\section*{Acknowledgements}
ATAW acknowledges with thanks support from Australian Research Council  grant DP220102232.

\bibliographystyle{biometrika}

\end{document}


\jname{Submitted to Biometrika}


\markboth{JIAZHEN XU et~al.}{Quantifying Periodicity in Non-Euclidean Object-Valued Time Series}

\title{Quantifying Periodicity in Non-Euclidean Object-Valued Time Series}

\author{JIAZHEN XU}
\affil{Department of Actuarial Studies and Business Analytics, Macquarie University, Sydney, 2113, Australia
\email{jiazhen.xu@mq.edu.au}}

\author{ANDREW T. A. WOOD}
\affil{Research School of Finance, Actuarial Studies and Statistics, Australian National University, Canberra ACT 2600, Australia
\email{andrew.wood@anu.edu.au}}

\author{\and TAO ZOU}
\affil{Research School of Finance, Actuarial Studies and Statistics, Australian National University, Canberra ACT 2600, Australia
\email{tao.zou@anu.edu.au}}
\maketitle

This Supplementary Material consists of five parts. \S\ref{supp sect::failure} illustrate limitations of extrinsic methods, \S\ref{supp sect::main result proof} covers the proofs of main theoretical results in the paper. \S\ref{sub sect::simulation sphere additional} contains the visualization of the simulated spherical time series. \S\ref{sub sect::simulation network} and \ref{sub sect::simulation function} covers additional simulations on networks and distributional data. \S\ref{supp sect:: real data figure} provides an additional figure relating to the real data analysis.

\makeatletter
\renewcommand\thesection{S\@arabic\c@section}
\renewcommand\thetable{S\@arabic\c@table}
\renewcommand \thefigure{S\@arabic\c@figure}
\makeatother

\section{Limitations of Extrinsic Methods}\label{supp sect::failure}

\cite{vogt2014nonparametric} focus on univariate time series. Here, we first extend the problem setup in \cite{vogt2014nonparametric} to multivariate cases. For real-valued time series $y_1,y_2,\ldots,y_T\in\mathbb{R}^p$ with a positive integer $p$, it is assumed that $y_t=m_{\rm E}(t)+\epsilon_t$ for a deterministic periodic function $m_{\rm E}(\cdot)$ with period $\theta_0$, where $E(\epsilon_t)=0$. Under this setting, we have $E(y_t)=m_{\rm E}(t)$ and one can observe that, if $m_{\rm E}(1)=m_{\rm E}(2)=\cdots=m_{\rm E}(\theta_0)$, the true period will become 1 and the method in \cite{vogt2014nonparametric} will gives the period estimation at 1.

Suppose we can embed object-valued time series $Y_1,Y_2,\cdots,Y_T$ into the Euclidean space $\mathbb{R}^p$. Then one can apply the method in \cite{vogt2014nonparametric}. Note that the idea of embedding non-Euclidean data into the Euclidean space, and then conducting Euclidean-based methods belongs to the class of extrinsic methods, compared to our proposed intrinsic method. A fundamental limitation of extrinsic methods is their failure to account for the intrinsic geometry of the underlying metric space. Below, we provide an example to illustrate this.

For $t=1,2,\ldots,T$, consider $Y_t \in \mathcal{S}^1 \subset \mathbb{R}^2$ which is represented as a vector parameterized by an angle $\varsigma \in (-\pi, \pi]$, that is $Y_t=\left( \cos(\varsigma), \sin(\varsigma) \right)^\top$. The data generating process is designed as a square wave with a true intrinsic period of $\theta_0 = 8$, governed by two alternating distributional regimes A and B. During the first four steps of the cycle, observations are drawn from regime A and for the subsequent four steps, observations are drawn from regime B.

Regime A is defined by three mass points $\varsigma \in \{0, 2\pi/3, -2\pi/3\}$, with respective probabilities $p = \{0.4, 0.3, 0.3\}$, while regime B is defined by mass points $\varsigma \in \{\pi, \pi/3, -\pi/3\}$, with respective probabilities $p = \{8/30, 11/30, 11/30\}$. For the extrinsic method, we directly embed the spherical data into $\mathbb{R}^2$ and calculations show that, for regime A where $t\in\{1,2,3,4\}$,
\begin{align*}
    m_{\rm E} (t) =  0.4 \begin{pmatrix} 1 \\ 0 \end{pmatrix} + 0.3 \begin{pmatrix} -1/2 \\ \sqrt{3}/2 \end{pmatrix} + 0.3 \begin{pmatrix} -1/2 \\ -\sqrt{3}/2 \end{pmatrix} = \begin{pmatrix} 0.1 \\ 0 \end{pmatrix},
\end{align*}
while regime B where $t\in\{5,6,7,8\}$,
\begin{align*}
    m_{\rm E} (t)  = \frac{8}{30} \begin{pmatrix} -1 \\ 0 \end{pmatrix} + \frac{11}{30} \begin{pmatrix} 1/2 \\ \sqrt{3}/2 \end{pmatrix} + \frac{11}{30} \begin{pmatrix} 1/2 \\ -\sqrt{3}/2 \end{pmatrix} = \begin{pmatrix} 0.1 \\ 0 \end{pmatrix}.
\end{align*}
This shows that the extrinsic method will lead to the issue caused by $m_{\rm E}(1)=m_{\rm E}(2)=\cdots=m_{\rm E}(\theta_0)$, because the two regimes become indistinguishable after Euclidean embedding.

In contrast, for our proposed intrinsic method where we consider the Fr\'{e}chet mean in \S\ref{subsection::problem setup}. Suppose the sphere $\mathcal{S}$ is equipped with the intrinsic metric, that is, the great circle distance. Then for regime A, the intrinsic Fr\'{e}chet mean minimizes the expected squared geodesic distance and by symmetry, the minimizer must be $\left( \cos(\varsigma), \sin(\varsigma) \right)^\top$ for either $\varsigma=0$ or $\varsigma=\pi$. Calculate $V(\nu) = E\{d^2(Y, \mu)\}$ for both candidates gives $V((1,0)^\top)=4\pi^2/15$ and $V((-1,0)^\top)=7\pi^2/15$, and thus $m(t)=(1,0)^\top$ for $t\in\{1,2,3,4\}$. Similarly, for regime B, symmetry dictates the Fr\'{e}chet mean must be $\left( \cos(\varsigma), \sin(\varsigma) \right)^\top$ for either $\varsigma=0$ or $\varsigma=\pi$. Calculate $V(\nu)$ for both candidates gives $V((1,0)^\top)=44\pi^2/135$ and $V((-1,0)^\top)=47\pi^2/135$, and thus $m(t)=(-1,0)^\top$ for $t\in\{5,6,7,8\}$. Thus we avoid the situation $m(1)=m(2)=\cdots=m(\theta_0)$ seen in the extrinsic case and still have the true period be 8.

\begin{figure}[!htbp]
\centering
\includegraphics[width=1\linewidth]{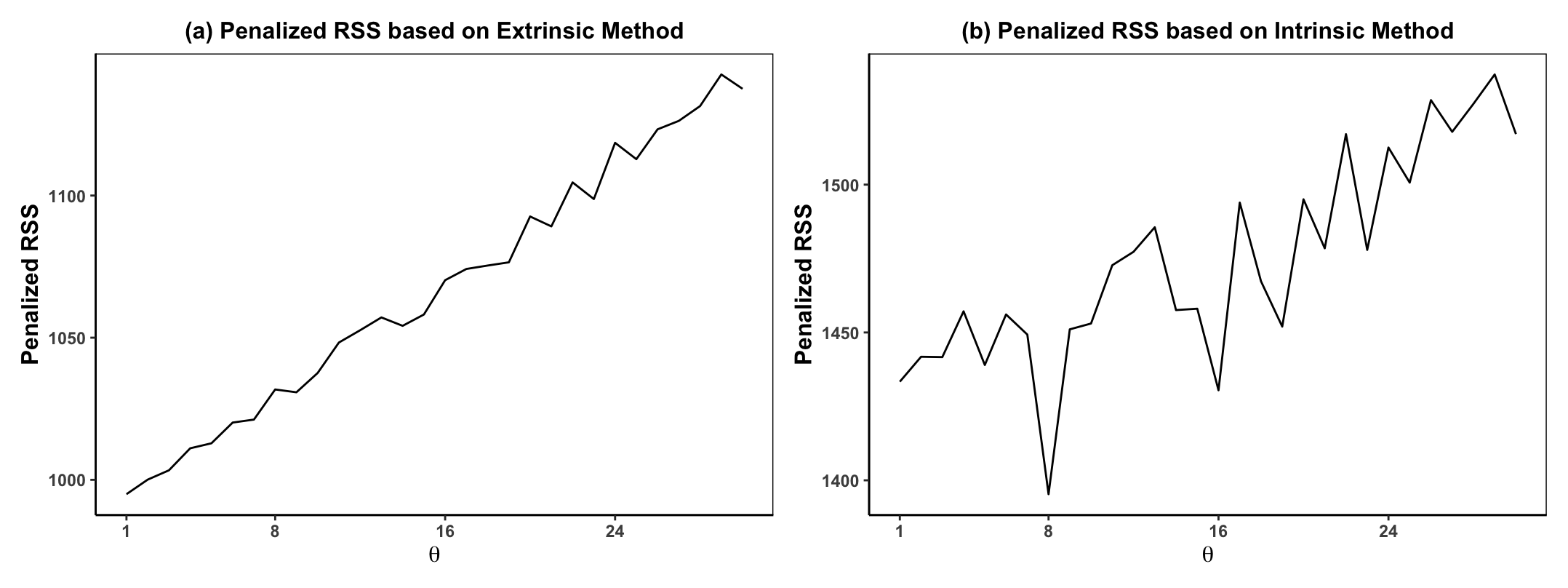}

\caption{ Plots of penalized RSS for one simulation replicate for periodic spherical time series with $T=2000$ using (a) the extrinsic method in (b) the intrinsic method using the information criterion for the residual sum of squares. \label{fig::extrinsic intrinsic compare}}
\end{figure}

Figure~\ref{fig::extrinsic intrinsic compare} visualizes the penalized RSS for the spherical time series generated via the data generating process described above. As expected, the extrinsic method leads to regimes A and B becoming indistinguishable, and the line in Fig.~\ref{fig::extrinsic intrinsic compare} (a) achieves the minimum value at 1. The intrinsic method used in Fig.~\ref{fig::extrinsic intrinsic compare} (b) gives the estimated period be 8, aligned with the true period $\theta_0$.

This example shows that, the extrinsic method which is built upon the Euclidean embedding may fail in some situations. Indeed, if the metric space $\Omega$ is a subset of the Euclidean space and the metric $d$ is chosen to be the induced Euclidean distance, our proposed method will be equivalent to the extrinsic method. This section also highlights the importance of the choice of metric.

\section{Proofs of Main Results}\label{supp sect::main result proof}

\begin{proof}[of Theorem~\ref{thm::uniform conv}]
    The proof consists of two parts. We first show the uniform convergence $\sup_{x\in\mathcal{X}}d(\widehat{\varkappa}(x),\varkappa(x))=o_p(1)$, and then use the chaining techniques to show the uniform convergence rate.

    \textbf{Part I.} In this part, we aim to show $\sup_{x\in\mathcal{X}}d(\widehat{\varkappa}(x),\varkappa(x))=o_p(1)$. Utilizing Theorem 2 in \cite{hong2026uniform}, it is sufficient to show that 
    \begin{align}\label{eq::uniform conv sufficient condition}
        \sup_{x\in\mathcal{X}}\sup_{\nu\in\Omega} \left| \widehat{\Xi}(\nu, x) -\Xi(\nu,x) \right| =o_p(1).
    \end{align}
    Recall that
    \begin{align*}
        \sup_{x\in\mathcal{X}}\sup_{\nu\in\Omega} \left| \widehat{\Xi}(\nu, x) -\Xi(\nu,x) \right| = \sup_{x\in\mathcal{X}}\sup_{\nu\in\Omega} \left| T^{-1}\sum_{t=1}^T w_t(x) \left[ d^2(Y_t,\nu) - E \{  d^2(Y_t,\nu) \} \right] \right|.
    \end{align*}
    Under Assumption~\ref{assumption 1}, $\{Y_t\}_{t\in\mathbb{Z}}$ is $\alpha$-mixing, then $\{d^2(Y_t,\nu)\}_{t\in\mathbb{Z}}$ is also $\alpha$-mixing with the mixing coefficients smaller than the ones of $\{Y_t\}_{t\in\mathbb{Z}}$. Moreover, let $Z_t=d^2(Y_t,\nu) - E \{  d^2(Y_t,\nu)\}$, recall that $\Omega$ is totally bounded and  $w_{\rm max}=\sup_{t\in\{1,2,\cdots,T\}}\sup_{x\in\mathcal{X}}|w_t(x)|<\infty$, one obtains
    \begin{align*}
        {\rm var} \left( T^{-1} \sum_{t=1}^T w_t(x) Z_t  \right) &= T^{-2}\sum_{t=1}^T w_t^2(x) {\rm var} (Z_t) + 2T^{-2}\sum_{1\leq i<j\leq T} w_i(x)w_j(x) {\rm cov}(Z_i,Z_j)\\
        &=O\left(T^{-2}\sum_{t=1}^T w_t^2(x)\right) =o(1)
    \end{align*}
    This gives the law of large numbers for the mixing sequence, that is
    \begin{align}\label{eq::unifrom conv pointwise conv}
    \left| T^{-1} \sum_{t=1}^T w_t(x) \left[ d^2(Y_t,\nu) - E \{  d^2(Y_t,\nu) \} \right] \right| = o_p(1).    
    \end{align}
    We now verify the asymptotically equicontinuity in probability, we aim to show that for any $\iota>0$, $\nu_1, \nu_2 \in \Omega$ and $x_1, x_2 \in \mathcal{X}$, 
    \begin{align}\label{formula::global conv equi cont}
         \limsup_T {\rm pr}\left( \sup_{d(\nu_1,\nu_2)<\zeta_1} \sup_{d_{\mathcal{X}}(x_1,x_2)<\zeta_2} \left| \overline{\Xi}(\nu_1,x_1)-  \overline{\Xi}(\nu_2,x_2) \right| >\iota \right)\to 0,
    \end{align}
    as certain sequences $\zeta_1,\zeta_2\to0$ where $\overline{\Xi}(\nu,x)= \widehat{\Xi}(\nu, x) -\Xi(\nu,x)$. Observe that
    \begin{align*}
        \left| \overline{\Xi}(\nu_1,x_1)-  \overline{\Xi}(\nu_2,x_2) \right| & \leq  \left| \overline{\Xi}(\nu_1,x_1)-  \overline{\Xi}(\nu_2,x_1) \right| +  \left| \overline{\Xi}(\nu_2,x_1)-  \overline{\Xi}(\nu_2,x_2) \right| \\
        & = O_p\left(d_{\mathcal{X}}(x_1 , x_2) \right) + O_p\left(d(\nu_1,\nu_2)\right).
    \end{align*}
    This implies that 
    \[
   \sup_{d(\nu_1,\nu_2)<\zeta_1} \sup_{d_{\mathcal{X}}(x_1,x_2)<\zeta_2} \left| \overline{\Xi}(\nu_1,x_1)-  \overline{\Xi}(\nu_2,x_2) \right| = O_p(\zeta_1) + O_p(\zeta_2),
    \]
    which proves (\ref{formula::global conv equi cont}). (\ref{eq::unifrom conv pointwise conv}) and (\ref{formula::global conv equi cont}) implies that (\ref{eq::uniform conv sufficient condition}) holds.

    \textbf{Part II.} In this part for the uniform convergence rate, we first develop a Bernstein-type inequality, and then utilize Theorem 2.2.19 in \cite{vanderVaartWellner2023} to obtain a maximal inequality, and finally, we apply the peeling technique to obtain the uniform convergence rate.

    \textbf{Part II (i).} Here we aim to analyse the centred empirical process $\mathbb{G}_T(\nu, x) = T^{-1}\sum_{t=1}^T Z_t(\nu, x)$ where
    \[
    Z_t(\nu, x) = w_t(x) \left\{ d^2(\nu, Y_t) - d^2(\varkappa(x), Y_t) \right)\} - E\left[ w_t(x) \left\{ d^2(\nu, Y_t) - d^2(\varkappa(x), Y_t) \right\} \right]
    \] 
    Let
    \begin{align*}
    \Delta_t(Y_t) =& w_t(x) \left\{ d^2(\nu, Y_t) - d^2(\varkappa(x), Y_t) \right\} - w_t(x^\prime) \left\{ d^2(\nu^\prime, Y_t) - d^2(\varkappa(x^\prime), Y_t) \right\}   \\
    =& w_t(x) \left\{ d^2(\nu, Y_t) - d^2(\nu^\prime, Y_t) - d^2(\varkappa(x), Y_t) + d^2(\varkappa(x^\prime), Y_t) \right\} \\
    &+ \left\{ w_t(x) - w_t(x^\prime) \right\} \left\{ d^2(\nu^\prime, Y_t) - d^2(\varkappa(x^\prime), Y_t) \right\}.
    \end{align*}
    Note that by the triangle inequality, we observe
    \begin{align*}
        |d^2(\nu, Y_t) - d^2(\nu^\prime, Y_t)| = |d(\nu, Y_t) + d(\nu^\prime, Y_t)||d(\nu, Y_t) - d(\nu^\prime, Y_t)| \leq 2 {\rm diam}(\Omega) d(\nu, \nu')    
    \end{align*}
    where ${\rm diam}(\Omega)$ is the diameter of the metric space $\Omega$. Similarly, one gets $|d^2(\nu^\prime, Y_t) - d^2(\varkappa(x^\prime), Y_t)| \leq 2{\rm diam}(\Omega) d(\nu^\prime, \varkappa(x^\prime)) \le 2\{{\rm diam}(\Omega)\}^2$ and $|d^2(\varkappa(x), Y_t) - d^2(\varkappa(x^\prime), Y_t)| \leq 2 {\rm diam}(\Omega) d(\varkappa(x), \varkappa(x^\prime))$. 
    Moreover, by Assumption~\ref{assumption 4}, we get $\Xi(x, \varkappa(x^\prime)) - \Xi(x, \varkappa(x)) \geq C_m d^\beta(\varkappa(x), \varkappa(x^\prime))$ and $\Xi(x^\prime, \varkappa(x)) - \Xi(x^\prime, \varkappa(x^\prime)) \geq C_m d^\beta(\varkappa(x), \varkappa(x^\prime))$. These gives
    \begin{align*}
        2 C_m d^\beta(\varkappa(x), \varkappa(x^\prime)) &\leq \Xi(x, \varkappa(x^\prime)) - \Xi(x, \varkappa(x)) + \Xi(x^\prime, \varkappa(x)) - \Xi(x^\prime, \varkappa(x^\prime)) \\
        & \leq 2 {\rm diam}(\Omega) L_w d_{\mathcal{X}}(x, x^\prime) d(\varkappa(x), \varkappa(x^\prime)),
    \end{align*}
    which shows 
    \[
    d(\varkappa(x), m(x')) \leq L_m d_{\mathcal{X}}(x, x')^{\frac{1}{\beta - 1}}
    \]
    for some positive $L_m$. These results imply that
    \begin{align}\label{ineq::incre bound}
        |\Delta_t(Y_t)| \leq 2 w_{\max} {\rm diam}(\Omega) \left[  d(\nu, \nu^\prime) +  L_m \{ d_{\mathcal{X}}(x, x')\}^{\frac{1}{\beta-1}} \right] + 2 L_w \{{\rm diam}(\Omega)\} ^2  d_{\mathcal{X}}(x, x^\prime) .
    \end{align}
    Let $\rho\left((\nu, x), (\nu^\prime, x^\prime)\right) = d(\nu, \nu^\prime) + \left\{d_{\mathcal{X}}(x, x^\prime)\right\}^{\frac{1}{\beta-1}} + d_{\mathcal{X}}(x, x^\prime)$. (\ref{ineq::incre bound}) gives the $L^\infty$ norm of $\overline{\Delta}_t=\Delta_t(Y_t) - E\{ \Delta_t(Y_t)\}$, $\|  \overline{\Delta}_t \|_\infty\leq M^*$ for $M^*=C_\Delta \rho\left((\nu, x), (\nu^\prime, x^\prime)\right)$ where $C_\Delta = 2 \max\{2{\rm diam}(\Omega) w_{\max}, 2{\rm diam}(\Omega) w_{\max} L_m, 2\{{\rm diam}(\Omega)\}^2 L_w\}$.

    Following the arguments in Part I, we also have
    \begin{align*}
       v^{*}=&\sup_{t=1,2,\ldots,T} \left[ {\rm var}\left( \overline{\Delta}_t  \right) + 2\sum_{j>t} \left| {\rm cov}\left(\overline{\Delta}_t, \overline{\Delta}_j\} \right)\right| \right] \\
       \leq & M^{*2} + 2 \sum_{k=1}^\infty 4\alpha(k) M^{*2} \leq C_1 w_{\max}^2 \rho^2\big((\nu, x), (\nu^\prime, x^\prime)\big)
    \end{align*}
    for some constant $C_1$ under Assumption~\ref{assumption 1}. Then, utilizing Theorem 2 in \cite{merlevede2009bernstein}, we obtain the Bernstein inequality
    \begin{align}\label{ineq::Bernstein}
        {\rm pr}\left( \left | \mathbb{G}_T(\nu, x) - \mathbb{G}_T(\nu^\prime, x^\prime) \right | > \epsilon \right) \leq \exp\left\{-\frac{C_2 T^2\epsilon^2}{v^{*2}T + M^{*2} + \epsilon M^* (\log T)^2 T}\right\}
    \end{align}
    for some constant $C_2$.

    \textbf{Part II (ii).} Note that the Bernstein inequality (\ref{ineq::Bernstein}) leads to the desired form on Page 153 in \cite{vanderVaartWellner2023} such as ${\rm pr}\left( \left | \mathbb{G}_T(\nu, x) - \mathbb{G}_T(\nu^\prime, x^\prime) \right | > \epsilon \right) \leq 2\exp \{ - \epsilon^2/(a+b \epsilon) \}$ where $a=(v^{*2}T + M^{*2})/C_2T^2$ and $b=M^* (\log T)^2/C_2 T$. Note that $h(\epsilon)=\frac{\epsilon^2}{2a} \mathbb{I}(\epsilon\leq a/b)+ \frac{\epsilon}{2b} \mathbb{I}(\epsilon> a/b) \leq \epsilon^2 /(a+b\epsilon)$ where $\mathbb{I}(\cdot)$ is the indicator function. Thus $\exp(-\epsilon^2 /(a+b\epsilon))\leq \exp(-h(\epsilon))$. Set $\exp(-h(\epsilon))=\exp(-x)$, as $h(\epsilon)$ is monotone and continuous, we get $h^{-1}(x)=\sqrt{2ax}\mathbb{I}(x\leq a/b^2) + 2bx \mathbb{I}(x> a/b^2) \leq \sqrt{2ax} + 2bx$. Thus we can conclude that (\ref{ineq::Bernstein}) implies
    \begin{align}\label{ineq::Bernstein 2}
    {\rm pr}\left( \left| \mathbb{G}_T(\nu, x) - \mathbb{G}_T(\nu^\prime, x^\prime) \right| > \sqrt{2ax} + 2bx \right) \le 2\exp(-x).
    \end{align}
    
    Let 
    \begin{align*}
        &\rho_1\left((\nu, x), (\nu^\prime, x^\prime)\right) = C_3 \frac{w_{\max} (\log T)^2 }{T}\rho\left((\nu, x), (\nu^\prime, x^\prime)\right) \\
        &\rho_2\left((\nu, x), (\nu^\prime, x^\prime)\right) = C_4 \frac{ w_{\max} }{\sqrt{T}}\rho\left((\nu, x), (\nu^\prime, x^\prime)\right)    
    \end{align*}
    for some constants $C_3$ and $C_4$. Then, utilizing Theorem 2.2.19 and Lemma 2.2.15 in \cite{vanderVaartWellner2023} and (\ref{ineq::Bernstein 2}) gives the maximal inequality
    \begin{align}\label{ineq::maximal ineq}
        E \left[ \sup_{(\nu, x) \in \mathcal{F}_\zeta} |\mathbb{G}_T(\nu, x)| \right] \leq &C_5 \int_0^{{\rm diam}(\mathcal{F}_\zeta)} \psi_1^{-1}(N(\epsilon, \mathcal{F}_\zeta, d_1)) d\epsilon \nonumber \\
        &+ C_6 \int_0^{{\rm diam}(\mathcal{F}_\zeta)} \psi_2^{-1}(N(\epsilon, \mathcal{F}_\zeta, d_2)) d\epsilon
    \end{align}
    for some constants $C_5$ and $C_6$, where $\mathcal{F}_\zeta = \{ (\nu, x) : x \in \mathcal{X}, d(\nu, \varkappa(x)) \leq \zeta \}$, $\psi_1^{-1}(x) = \log(1+x)$, and $\psi_2^{-1}(x) = \sqrt{\log(1+x)}$. 
    
    Recall that $N(r, \mathcal{X}, d_{\mathcal{X}}) = O(e(T))$ for any $r>0$, thus $N(\epsilon, \mathcal{F}_\zeta, \rho) \leq C_7 e(T) \sup_{x \in \mathcal{X}_T} N(\epsilon, B_\zeta(\varkappa(x)), d)$ for some constant $C_7$. Note that the second term dominates the first term in the right side of (\ref{ineq::maximal ineq}), and using a change of variables $u = \sqrt{T}\epsilon/\{C_4 w_{\max}\}$, we obtain, under Assumption~\ref{assumption 5},
    \begin{align*}
        &\int_0^{{\rm diam}(\mathcal{F}_\zeta)} \psi_2^{-1}(N(\epsilon, \mathcal{F}_\zeta, d_2)) d\epsilon\\
        &\leq \frac{ C_4 w_{\max}}{\sqrt{T}} \int_0^\zeta \sqrt{ \log e(T) + \sup_{x \in \mathcal{X}_T} \log N(u, B_\zeta(\varkappa(x)), d) }  du\\
        &=O\left( \frac{w_{\max} \sqrt{\log e(T)}}{\sqrt{T}} \zeta \right).
    \end{align*}
    
    
    The above analysis gives 
    \begin{align}\label{tail moment bound}
    E \left[ \sup_{(\nu, x) \in \mathcal{F}_\zeta} |\mathbb{G}_T(\nu, x)| \right]  = O\left( \frac{w_{\max} \sqrt{\log e(T)}}{\sqrt{T}} \zeta^\gamma \right) = O\left( \frac{\sqrt{\log e(T)}\zeta^\gamma}{\sqrt{T}} \right)  
    \end{align}
    for $\gamma \leq 1$, where the last equality holds as $w_{\rm max}=O(1)$.

    \textbf{Part II (iii).} Here we apply the peeling technique to obtain the uniform convergence rate. Consider $\alpha_T=(T/\log e(T))^{\beta^\prime/\{4(\beta^\prime-1)\}}$ for any $\beta^\prime > \beta$ and define
    \[
    \mathcal{C}_{l,T}=\{(\nu,x):x\in\mathcal{X},2^{l-1}< \alpha_T d^{\beta^\prime/2}(\nu,\varkappa(x)) \leq 2^l \}.
    \]
    Under Assumption~\ref{assumption 4}, for any integer $N$, there exists some constants $C_8$ and $C_9$ such that
    \begin{align*}
        {\rm pr}\left( \sup_{x\in\mathcal{X}} \alpha_T d^{\beta^\prime/2} (\widehat{\varkappa}(x),\varkappa(x)) >2^N \right)  
        &\leq  {\rm pr}(\sup_{x\in\mathcal{X}}2d(\widehat{\varkappa}(x),\varkappa(x))>C_8) \\
        &+ \sum_{l \geq N} {\rm pr}\left( \sup_{(\nu, x) \in \mathcal{C}_{l,T}} |\mathbb{G}_T(\nu, x)| \geq C_9 \zeta_{l-1}^\beta \right),
    \end{align*}
    where $\zeta_l = \left( \frac{2^l}{\alpha_T} \right)^{2/\beta'}$. By Part I, $ {\rm pr}(\sup_{x\in\mathcal{X}}2d(\widehat{\varkappa}(x),\varkappa(x))>C_7) \to 0$ as $T\to\infty$. Moreover, (\ref{tail moment bound}) shows that 
    \begin{align*}
    &\sum_{l \geq N} {\rm pr}\left( \sup_{(\nu, x) \in \mathcal{C}_{l,T}} |\mathbb{G}_T(\nu, x)| \geq C_9 \zeta_{l-1}^\beta \right) \leq C_9 \sum_{l\geq N} \frac{\sqrt{\log e(T)}\zeta_l^\gamma}{\sqrt{T}\zeta_{l-1}^\beta}\\
    &= C_{10} \sum_{l\geq N}\frac{2^{2\beta/\beta^\prime} \sqrt{\log e(T)}}{\sqrt{T}} (2^l)^{ \frac{2(\gamma - \beta)}{\beta^\prime} } (\alpha_T)^{ -\frac{2(\gamma - \beta)}{\beta^\prime} } 
    \leq  C_{11} \sum_{l \geq N} \left( \frac{1}{4^{(\beta^\prime - 1)/\beta^\prime}} \right)^l
    \end{align*}
    for some positive constants $C_{10}$ and $C_{11}$, where the last inequality holds by choosing $\gamma = 1 + \beta - \beta^\prime$. Note the right-hand side converges as $\beta^\prime>1$. Therefore, by choosing a sufficient large $N$, we have 
    \[
    \sup_{x\in\mathcal{X}}d(\widehat{\varkappa}(x),\varkappa(x)) = O_p\left(\alpha_n^{-2/\beta^\prime}\right)=O_p\left(\left(\frac{T}{\log e(T)}\right)^{-1/\{2(\beta^\prime-1)\}}\right).
    \]
    This completes the proof.
\end{proof}

\begin{proof}[of Proposition~\ref{prop::uniform convergence rate partial mean}]
    We first evaluate the covering number of $\mathcal{X}$ with the metric $d_{\mathcal{X}}$ be the Euclidean metric. Recall that $\mathcal{X}$ is a discrete space of mutually orthogonal coordinate vectors, thus its cardinality is ${\rm card}(\mathcal{X}) = \Theta_T(\Theta_T+1)/2 = O(\Theta_T^2)$. Moreover, notice that the distance between any two distinct vectors in $\mathcal{X}$ is $\sqrt{2}$, the covering number $N(r, \mathcal{X}, d_{\mathcal{X}})$ behaves as $N(r, \mathcal{X}, d_{\mathcal{X}})={\rm card}(\mathcal{X})$ for any radius $r < \sqrt{2}/2$ as every point needs a ball to cover. When $r \geq \sqrt{2}$, one ball can cover multiple points. Therefore, for any radius $r > 0$, the covering number is strictly bounded by the cardinality of the set, that is $N(r, \mathcal{X}, d_{\mathcal{X}}) \leq {\rm card}(\mathcal{X}) = O(\Theta_T^2)$. 

    Observe that based on the construction, $w_{\rm max}=\sup_{t\in\{1,2,\cdots,T\}}\sup_{x\in\mathcal{X}}|w_t(x)|\leq 1$. Moreover, we have
    \[
    |w_t(x) - w_t(x^\prime)| \leq d_{\mathcal{X}}(x,x^\prime).
    \]
    Finally, note that (\ref{formula::specific Frechet regression}) gives the effective size being $T/\Theta_T$. Under the assumption that $(\Theta_T\log\Theta_T)/T \to 0$ as $T\to\infty$, and then applying Theorem~\ref{thm::uniform conv}, we have
    \[
    \sup_{x\in\mathcal{X}}d(\widehat{\varkappa}(x),\varkappa(x)) =O_p\left(\left(\frac{T}{\Theta_T\log \Theta_T}\right)^{-1/\{2(\beta^\prime-1)\}}\right)
    \]
    
\end{proof}

\begin{proof}[of Theorem \ref{thm::period estimation consistency}]

    Recall $M(\theta) = \lfloor T/\theta\rfloor$ and  $n(t,\theta)=M(\theta)+\mathbb{I}(r(t,\theta)\leq r(T,\theta))$ for any $\theta$ such that $1\leq \theta \leq \Theta_T$ with $\Theta_T/T\to 0$ as $T\to\infty$, $\lfloor\cdot\rfloor$ being the floor function, $\mathbb{I}(\cdot)$ being the indicator function, and $r(t,\theta)=t+\theta-\theta\lfloor (t+\theta-1)/\theta \rfloor$. For observations $\{Y_{r(t,\theta)+(i-1)\theta}\}_{i=1,2,\ldots,n(t,\theta)}$, denote the corresponding population Fr\'{e}chet mean as $\widetilde{\mu}_{t,\theta}$. 
    
    Here, we first show that under Assumptions~\ref{assumption 1}--\ref{assumption 5},
    \begin{align}\label{statement_of_Lemma_S3}
    \sup_{1\leq\theta\leq \Theta_T}T^{-1}{\rm RSS}(\theta) = &\sup_{1\leq\theta\leq \Theta_T} \left[\sum_{l=1}^\theta w_l E\{d^2(Y_l,\widetilde{\mu}_{l,\theta})\} \right] \nonumber\\
    &+O_p\left(\min\left\{ (T/\Theta_T)^{-1/2},\{T/(\Theta_T\log \Theta_T)\}^{-1/(\beta^\prime-1)} \right\}\right) 
    \end{align}
    hold for some positive weights $w_l,l=1,2,\ldots,\theta$ with $\sum_{l=1}^{\theta}w_l=1$, and any $\beta^\prime>\beta$ with $\beta$ given in Assumption~\ref{assumption 4}. To achieve this, we first establish that the right-hand side of (\ref{statement_of_Lemma_S3}) is an upper bound for $T^{-1}{\rm RSS}(\theta)$ and then we establish that it is also a lower bound.

    Observe that, by the triangle inequality, we have
    \begin{align*}
        T^{-1}{\rm RSS}(\theta)  \leq T^{-1}\sum_{t=1}^T \left[ d(Y_t,\widetilde{\mu}_{t,\theta}) + d(\widehat{m}_t (\theta,T),\widetilde{\mu}_{t,\theta}) \right]^2= I + II + III,
    \end{align*}
    where 
    \[
    I = T^{-1}\sum_{t=1}^T d^2(Y_t,\widetilde{\mu}_{t,\theta}),~~II = T^{-1}\sum_{t=1}^T d^2(\widehat{m}_t (\theta,T),\widetilde{\mu}_{t,\theta}),
    \]
    and 
    \[
    III = T^{-1}\sum_{t=1}^T 2d(Y_t,\widetilde{\mu}_{t,\theta})d(\widehat{m}_t (\theta,T),\widetilde{\mu}_{t,\theta}).
    \]
    By the law of large numbers for the mixing sequence, see the Part I of the proof of Theorem~\ref{thm::uniform conv}, one can see that 
    \begin{align}\label{eq::lemma I asy}
        \sup_{1\leq\theta\leq \Theta_T}I = \sup_{1\leq\theta\leq \Theta_T} \left[\sum_{l=1}^\theta w_l E\{d^2(Y_l,\widetilde{\mu}_{l,\theta})\}\right] + O_p\left\{(T/\Theta_T)^{-1/2}\right\},
    \end{align}
    as the effect sample size is $T/\Theta_T$. By the uniform convergence rate given in Proposition~\ref{prop::uniform convergence rate partial mean}, we have 
    \begin{align}\label{eq::uniform conv rate}
        \sup_{\substack{1\leq \theta\leq\Theta_T\\1\leq t\leq T}}d(\widehat{m}_t (\theta,T),\widetilde{\mu}_{t,\theta})=\sup_{x\in\mathcal{X}}d(\widehat{\varkappa}(x),\varkappa(x)) =O_p\{(T/\log \Theta_T)^{-1/2(\beta^\prime-1)}\} .
    \end{align}
    To see why the first equality holds, note that by our construction, $\widetilde{\mu}_{t,\theta}=m_t(\theta, T)$, 
    \[
    \left\{ d\left(\widehat{\varkappa}(x),\varkappa(x)\right) : x \in \mathcal{X} \right\} = \left\{ d\left(\widehat{m}_t (\theta,T),m_t(\theta, T)\right) : 1\leq \theta\leq\Theta_T, \, 1\leq t\leq T \right\},
    \]
    and observe that taking the supremum over identical sets naturally yields identical values.
    
    (\ref{eq::uniform conv rate}) gives
    \begin{align}\label{eq::lemma II asy}
       \sup_{1\leq\theta\leq \Theta_T} II = O_p\left\{\{T/(\Theta_T\log \Theta_T)\}^{-1/(\beta^\prime-1)}\right\}.
    \end{align}
    Moreover, observe that
    \begin{align*}
    III = & T^{-1}\sum_{t=1}^T 2\left\{d(Y_t,\widetilde{\mu}_{t,\theta}) - E[d(Y_t,\widetilde{\mu}_{t,\theta})]\right\}d(\widehat{m}_t (\theta,T),\widetilde{\mu}_{t,\theta}) \\
    & + T^{-1}\sum_{t=1}^T 2E[d(Y_t,\widetilde{\mu}_{t,\theta})]d(\widehat{m}_t (\theta,T),\widetilde{\mu}_{t,\theta}) \\.
    \end{align*}
    Then following similar techniques used in deriving (\ref{eq::lemma I asy}) and (\ref{eq::lemma II asy}), we get
    \begin{align}\label{eq::lemma III asy}
    \sup_{1\leq\theta\leq \Theta_T}III =  O_p\left\{(T/\Theta_T)^{-1/2}\{T/(\Theta_T\log \Theta_T)\}^{-1/(\beta^\prime-1)}\right\}.
    \end{align}
    Combining (\ref{eq::lemma I asy}), (\ref{eq::lemma II asy}) and (\ref{eq::lemma III asy}) gives that
    \begin{align*}
        \sup_{1\leq\theta\leq \Theta_T}T^{-1}{\rm RSS}(\theta) \leq &\sup_{1\leq\theta\leq \Theta_T} \left[\sum_{l=1}^\theta w_l E\{d^2(Y_l,\widetilde{\mu}_{l,\theta})\} \right] \\
    &+O_p\left(\min\left\{ (T/\Theta_T)^{-1/2},\{T/(\Theta_T\log \Theta_T)\}^{-1/(\beta^\prime-1)} \right\}\right) 
    \end{align*}
    This completes the proof of the upper bound.

    For the lower bound, note that by the triangle inequality, we also have
    \begin{align*}
        T^{-1}{\rm RSS}(\theta) & \geq T^{-1}\sum_{t=1}^T \left[ d(Y_t,\widetilde{\mu}_{t,\theta}) - d(\widehat{m}_t (\theta,T),\widetilde{\mu}_{t,\theta}) \right]^2\\
        &= I + II - III.
    \end{align*}
    By (\ref{eq::lemma I asy}), (\ref{eq::lemma II asy}) and (\ref{eq::lemma III asy}), we can conclude that 
    \begin{align*}
        \sup_{1\leq\theta\leq \Theta_T}T^{-1}{\rm RSS}(\theta) \geq &\sup_{1\leq\theta\leq \Theta_T} \left[\sum_{l=1}^\theta w_l E\{d^2(Y_l,\widetilde{\mu}_{l,\theta})\} \right] \\
    &+O_p\left(\min\left\{ (T/\Theta_T)^{-1/2},\{T/(\Theta_T\log \Theta_T)\}^{-1/(\beta^\prime-1)} \right\}\right) ,
    \end{align*}
    and thus (\ref{statement_of_Lemma_S3}) holds.

    Denote $\mathcal{M}_{\theta_0}=\{\theta: \theta = k\theta_0,k=1,2,\ldots,1\leq \theta \leq \Theta_T\}$. To prove Theorem~\ref{thm::period estimation consistency}, we only need to prove the following two results:
    \begin{align}
        &{\rm pr}\left( \min_{\theta\in\mathcal{M}_{\theta_0},\theta\neq \theta_0} \mathcal{L}(\theta,\lambda_T) > \mathcal{L}(\theta_0,\lambda_T) \right) \to 1, \label{formula::thm1 goal1}\\
        &{\rm pr}\left( \min_{\theta\notin\mathcal{M}_{\theta_0},1\leq \theta\leq \Theta_T} \mathcal{L}(\theta,\lambda_T) > \mathcal{L}(\theta_0,\lambda_T) \right) \to 1\label{formula::thm1 goal2}, 
    \end{align}
    as $T\to\infty$.

    For the first case where $\theta\in\mathcal{M}_{\theta_0},\theta\neq \theta_0$. Note that under the first case, $\theta=K\theta_0$ for some integer $K>1$. Recall that $\widetilde{\mu}_{t,\theta}$ is the population Fr\'{e}chet mean of $\{Y_{r(t,\theta)+(i-1)\theta}\}_{i=1,2,\ldots,n(t,\theta)}$. Thus we get
    \begin{align*}
        &T^{-1}{\rm RSS}(\theta) - T^{-1}{\rm RSS}(\theta_0)\nonumber\\
        \geq& T^{-1}\sum_{t=1}^T \Big[ d^2(\widehat{m}_t(\theta,T),\mu_{r(t,\theta_0)} ) - d^2(\widehat{m}_t(\theta_0,T),\mu_{r(t,\theta_0)} ) \nonumber\\
        &- 2d(Y_t,\mu_{r(t,\theta_0)})d(\widehat{m}_t(\theta,T),\mu_{r(t,\theta_0)} ) - 2d(Y_t,\mu_{r(t,\theta_0)})d(\widehat{m}_t(\theta_0,T),\mu_{r(t,\theta_0)} ) \Big] ,
    \end{align*}
    and 
    \begin{align*}
        &T^{-1}{\rm RSS}(\theta) - T^{-1}{\rm RSS}(\theta_0)\nonumber\\
        \leq& T^{-1}\sum_{t=1}^T \Big[ d^2(\widehat{m}_t(\theta,T),\mu_{r(t,\theta_0)} ) - d^2(\widehat{m}_t(\theta_0,T),\mu_{r(t,\theta_0)} ) \nonumber\\
        &+ 2d(Y_t,\mu_{r(t,\theta_0)})d(\widehat{m}_t(\theta,T),\mu_{r(t,\theta_0)} ) + 2d(Y_t,\mu_{r(t,\theta_0)})d(\widehat{m}_t(\theta_0,T),\mu_{r(t,\theta_0)} ) \Big] .
    \end{align*}
    Then following the techniques used in the verification of (\ref{statement_of_Lemma_S3}) and noticing that $T^{-1}{\rm RSS}(\theta)\geq 0$ for any $\theta\in\{1,2,\ldots,\Theta_T\}$, we have
    \begin{align*}
        \min_{\theta\in\mathcal{M}_{\theta_0},\theta\neq \theta_0} \left\{T^{-1}{\rm RSS}(\theta) - T^{-1}{\rm RSS}(\theta_0) \right\} = O_p\left[\{T/(\Theta_T\log \Theta_T)\}^{-1/(\beta^\prime-1)}\right],
    \end{align*}
     This result gives
    \begin{align*}
        \min_{\theta\in\mathcal{M}_{\theta_0},\theta\neq \theta_0}T^{-1}\left[\mathcal{L}(\theta,\lambda_T) - \mathcal{L}(\theta_0,\lambda_T) \right]
        = O_p\left[\{T/(\Theta_T\log \Theta_T)\}^{-1/(\beta^\prime-1)}\right] + T^{-1}\lambda_T(k-1)\theta_0.
    \end{align*}
    Note that when $T^{1-1/(\beta^\prime-1)}(\Theta_T\log \Theta_T)^{1/(\beta^\prime-1)}\ll \lambda_T \ll T$, the difference between the penalties in the loss functions $\mathcal{L}(\theta,\lambda_T)$ and $\mathcal{L}(\theta_0,\lambda_T)$ is the leading term uniformly over the set $\mathcal{M}_{\theta_0}$ excluding $\{\theta_0\}$. Denote such the set $\mathcal{M}_{\theta_0}$ excluding $\{\theta_0\}$ as $\mathcal{M}_{\theta_0} / \{\theta_0\}$. We have for sufficiently large $T$, $   \min_{\theta\in\mathcal{M}_{\theta_0} / \{\theta_0\}}T^{-1}\left[\mathcal{L}(\theta,\lambda_T) - \mathcal{L}(\theta_0,\lambda_T)\right] > 0$ with probability approaching 1, which indicates that (\ref{formula::thm1 goal1}) holds.

    We now consider the second case where $\theta\notin\mathcal{M}_{\theta_0},1\leq\theta\leq \Theta_T$.  One can see that $\widetilde{\mu}_{t,\theta}\neq \mu_{r(t,\theta_0)}$ for $t=1,\ldots,T$ when $\theta\notin\mathcal{M}_{\theta_0},1\leq\theta\leq \Theta_T$. 
    By (\ref{statement_of_Lemma_S3}), one has
    \begin{align}\label{ineq::thm1 MSE false period}
        &\min_{\theta\notin\mathcal{M}_{\theta_0}} \left\{T^{-1}{\rm RSS}(\theta) - T^{-1}{\rm RSS}(\theta_0)\right\}\nonumber\\
        =& \min_{\theta\notin\mathcal{M}_{\theta_0}} \sum_{l=1}^{\theta\theta_0}w_l\left\{E[d^2(Y_l,\widetilde{\mu}_{l,\theta})] - E[d^2(Y_l,\mu_{r(l,\theta_0)})] \right\}  + o_p(1)
    \end{align}
    for some positive weights $w_l,l=1,2,\ldots,\theta\theta_0$ with $\sum_{l=1}^{\theta\theta_0}w_l=1$. Notably, for $l=1,2,\ldots,\theta\theta_0$, $E\{d^2(Y_l,\mu_{r(t,\theta_0)})\} =\min_{\nu\in\Omega}E\{d^2(Y_l,\nu)\} $, which implies that $E\{d^2(Y_l,\mu_{r(t,\theta_0)})\}  < E\{d^2(Y_l,\widetilde{\mu}_{t,\theta})\}$. This result, together with (\ref{ineq::thm1 MSE false period}), imply that $T^{-1}\left[\mathcal{L}(\theta) - \mathcal{L}(\theta_0)\right]>0$ when $\theta\notin\mathcal{M}_{\theta_0},1\leq\theta\leq \Theta_T$. Thus we can conclude that, for sufficiently large $T$, $\min_{\theta\notin\mathcal{M}_{\theta_0}}T^{-1}\left[\mathcal{L}(\theta,\lambda_T) - \mathcal{L}(\theta_0,\lambda_T)\right] > 0$ with probability approaching 1, which indicates that (\ref{formula::thm1 goal2}) holds. This completes the proof.
\end{proof}

\begin{proof}[of Theorem \ref{thm::IC}]
    The proof consists of two parts, where we first consider ${\rm IC}_\lambda = \log \{ {\rm RSS}(\widehat{\theta}_\lambda) / T \} + \widehat{\theta}_\lambda g(T)$ and then study the behaviour of ${\rm IC}_\lambda = {\rm RSS}(\widehat{\theta}_\lambda) / T  + \widehat{\theta}_\lambda g(T)$.

    \textbf{Part I}: When ${\rm IC}_\lambda = \log \{ {\rm RSS}(\widehat{\theta}_\lambda) / T \} + \widehat{\theta}_\lambda g(T)$, we need to show that for any $\lambda\in \Lambda_+ \cup \Lambda_-$, ${\rm IC}_\lambda > {\rm IC}_{\lambda_T}$ with probability going to 1. Here we subsequently consider two cases when $\lambda\in \Lambda_+$ and $\lambda\in \Lambda_-$.

    \textbf{Case 1 of Part I}: We first consider the case when $\lambda\in \Lambda_+$. Recall that by Theorem \ref{thm::period estimation consistency}, we have $\widehat{\theta}_{\lambda_T} = \theta_0+o_p(1)$. Thus $\widehat{\theta}_\lambda - \widehat{\theta}_{\lambda_T} = (k-1)\theta_0$ for some integer $k>1$ with probability tending to 1. This gives that, with probability tending to 1,
    \begin{align}\label{ineq::thm2 case 1 BIC difference}
        \min_{\lambda\in\Lambda_+} \left( {\rm IC}_\lambda - {\rm IC}_{\lambda_T}\right) & =  \min_{\lambda\in\Lambda_+} \log\left\{ \frac{T^{-1}{\rm RSS}(\widehat{\theta}_\lambda)}{ T^{-1}{\rm RSS}(\widehat{\theta}_{\lambda_T})}\right\} + (k-1)\theta_0 g(T) .
    \end{align}
    By (\ref{statement_of_Lemma_S3}), one can see that 
    \begin{align}\label{eq::thm2 case 1 conv ip}
        T^{-1}{\rm RSS}(\widehat{\theta}_{\lambda_T})=C_1+o_p(1)
    \end{align}
    for some positive constant $C_1$. Therefore, after utilizing Slutsky's Lemma and then following the techniques used in the proof of Theorem \ref{thm::period estimation consistency}, based on the definition of $\Lambda_+$, we get
    \begin{align}\label{eq::thm2 case 1 division ip}
        \min_{\lambda\in \Lambda_+}\frac{T^{-1}{\rm RSS}(\widehat{\theta}_\lambda) - T^{-1}{\rm RSS}(\widehat{\theta}_{\lambda_T})}{  T^{-1}{\rm RSS}(\widehat{\theta}_{\lambda_T})}  = O_p\left[\{T/(\Theta_T\log \Theta_T)\}^{-1/(\beta^\prime-1)}\right].
    \end{align}
    This result, together with the Taylor series expansion, gives
    \begin{align}\label{eq::thm2 case 1 log term Taylor}
        &\min_{\lambda\in \Lambda_+}\log\left\{ \frac{T^{-1}{\rm RSS}(\widehat{\theta}_\lambda)}{ T^{-1}{\rm RSS}(\widehat{\theta}_{\lambda_T})}\right\}
        =  \min_{\lambda\in \Lambda_+}\log\left\{ 1+ \frac{T^{-1}{\rm RSS}(\widehat{\theta}_\lambda) - T^{-1}{\rm RSS}(\widehat{\theta}_{\lambda_T})}{  T^{-1}{\rm RSS}(\widehat{\theta}_{\lambda_T})}\right\} \nonumber\\
         = & \min_{\lambda\in \Lambda_+}\frac{T^{-1}{\rm RSS}(\widehat{\theta}_\lambda) - T^{-1}{\rm RSS}(\widehat{\theta}_{\lambda_T})}{  T^{-1}{\rm RSS}(\widehat{\theta}_{\lambda_T})} + o_p\left(\min_{\lambda\in \Lambda_+}\frac{T^{-1}{\rm RSS}(\widehat{\theta}_\lambda) - T^{-1}{\rm RSS}(\widehat{\theta}_{\lambda_T})}{  T^{-1}{\rm RSS}(\widehat{\theta}_{\lambda_T})}\right)\nonumber\\
         = & O_p\left[\{T/(\Theta_T\log \Theta_T)\}^{-1/(\beta^\prime-1)}\right],
    \end{align}
    where the last equality holds by (\ref{eq::thm2 case 1 division ip}). After borrowing the techniques used in the proof of Theorem \ref{thm::period estimation consistency}, together with (\ref{ineq::thm2 case 1 BIC difference}) and (\ref{eq::thm2 case 1 log term Taylor}), the proof under this case is completed by observing $(k-1)\theta_0 g(T)$ is the leading term under the condition (i) where $g(T)\gg \{T/(\Theta_T\log \Theta_T)\}^{-1/(\beta^\prime-1)}$.

    \textbf{Case 2 of Part I}: We then consider the case when $\lambda\in \Lambda_-$. By (\ref{eq::thm2 case 1 conv ip}), one can see that
    \begin{align*}
          \min_{\lambda\in \Lambda_-} \log\left\{ \frac{{\rm RSS}(\widehat{\theta}_\lambda)}{ {\rm RSS}(\widehat{\theta}_{\lambda_T})}\right\}  &=\min_{\lambda\in \Lambda_-}  \log\left\{ 1+ \frac{T^{-1}{\rm RSS}(\widehat{\theta}_\lambda) - T^{-1}{\rm RSS}(\widehat{\theta}_{\lambda_T})}{  T^{-1}{\rm RSS}(\widehat{\theta}_{\lambda_T})}\right\} \\
          &=\log(1+C_2) + o_p(1),
    \end{align*}
    for some positive constant $C_2$, where the final equality holds by following the techniques used in the proof of Theorem \ref{thm::period estimation consistency}. Therefore, given the condition (ii) where $g(T)=o(1)$, we have
    \[
    \sup_{\lambda\in \Lambda_-}  \left({\rm IC}_\lambda - {\rm IC}_{\lambda_T}  \right)= C_5 + o_p(1).
    \]
    for some positive constant $C_3$. This completes the proof in Case 2.

    \textbf{Part II}: We now consider ${\rm IC}_\lambda = {\rm RSS}(\widehat{\theta}_\lambda) / T  + \widehat{\theta}_\lambda g(T)$. Similar to Part I, we need to consider two cases when $\lambda\in \Lambda_+$ and $\lambda\in \Lambda_-$. 

    \textbf{Case 1 of Part II}: We first consider the case when $\lambda\in \Lambda_+$. Similar to Case 1 of Part I, we have 
    \[
    \min_{\lambda\in\Lambda_+}\left({\rm IC}_\lambda - {\rm IC}_{\lambda_T}\right)  \geq \min_{\lambda\in\Lambda_+} \left\{ T^{-1}{\rm RSS}(\widehat{\theta}_\lambda)- T^{-1}{\rm RSS}(\widehat{\theta}_{\lambda_T})\right\}  + (k-1)\theta_0 g(T),
    \]
    and 
    \[
    \min_{\lambda\in\Lambda_+} T^{-1}\left\{{\rm RSS}(\widehat{\theta}_\lambda)- {\rm RSS}(\widehat{\theta}_{\lambda_T}) \right\}= O_p\left[\{T/(\Theta_T\log \Theta_T)\}^{-1/(\beta^\prime-1)}\right].
    \]
    Note that $(k-1)\theta_0 g(T)$ is the leading term under the condition (i) where  $g(T)\gg \{T/(\Theta_T\log \Theta_T)\}^{-1/(\beta^\prime-1)}$, which completes the proof of this case.

    \textbf{Case 2 of Part II}: Finally, we consider the case when $\lambda\in \Lambda_-$. Borrowing the arguments used in the proof of Case 2 of Part I, we get
    \[
    \min_{\lambda\in\Lambda_-}  T^{-1}\left\{{\rm RSS}(\widehat{\theta}_\lambda)- {\rm RSS}(\widehat{\theta}_{\lambda_T}) \right\}= C_6 + o_p(1)
    \]
    for some positive constant $C_6$. Thus by the condition (ii) where $g(T)=o(1)$, one can see that ${\rm IC}_\lambda - {\rm IC}_{\lambda_T}  = C_6 + o_p(1)$. This completes the proof under this case
    
    Combining Parts I and II completes the proof of Theorem~\ref{thm::IC}.
\end{proof}

\begin{proof}[of Corollary \ref{cor::final period est}]
    Note that Theorem~\ref{thm::period estimation consistency} gives $\widehat{\theta}_{\lambda_T}=\theta_0+o_p(1)$,
    as $T\to\infty$ when $\lambda_T$ satisfies the condition in Theorem \ref{thm::period estimation consistency}. Thus by Theorem \ref{thm::IC}, if $\lambda$ falls into the set $\Lambda_+$ or $\Lambda_-$, then the corresponding ${\rm IC}_\lambda$ will be larger than the reference level ${\rm IC}_{\lambda_T}$ with probability tending to 1 as $T\to\infty$. This is achieved by (\ref{eq::thm IC statement}).

    Moreover, as $\widehat{\lambda}=\argmin_\lambda {\rm IC}_\lambda$, we always have ${\rm IC}_{\widehat{\lambda}}\leq {\rm IC}_{\lambda_T}$, which can ensure that $\widehat{\lambda}$ falls into the set $\Lambda_0=\{\lambda:\widehat{\theta}_\lambda=\theta_0\}$. This completes the proof.
\end{proof}

\begin{proof}[Proof of Theorem \ref{thm::periodic component est consistency}:]
    The aim of the proof is to show that for any $\eta>0$, there exists a constant $C=C(\eta)$ such that
    \begin{align}\label{ineq::thm component estimator goal}
        {\rm pr}\left( \max_{t=1,\ldots,T} d(\widehat{m}(t),m(t)) >\frac{C}{T^{-1/2(\beta^\prime-1)} } \right)<\eta
    \end{align}
    for sufficiently large $T$. To see this, note that
    \begin{align}\label{ineq::thm component estimator upper bound}
        &{\rm pr}\left( \max_{t=1,\ldots,T} d(\widehat{m}(t),m(t)) >\frac{C}{T^{-1/2(\beta^\prime-1)} } \right) \nonumber\\
        \leq & {\rm pr}\left( \max_{t=1,\ldots,T} d(\widehat{m}(t),m(t)) >\frac{C}{T^{-1/2(\beta^\prime-1)} } , \widehat{\theta}_{\widehat{\lambda}} = \theta_0\right) + {\rm pr}(\widehat{\theta}_{\widehat{\lambda}}\neq \theta_0).
    \end{align}
    By Theorems~\ref{thm::period estimation consistency} and \ref{thm::IC}, we have \begin{align}\label{ineq::thm component estimator ess1}
        {\rm pr}(\widehat{\theta}_{\widehat{\lambda}} \neq \theta_0)=o(1)
    \end{align}
    as $n\to\infty$. Let $\widetilde{m}(t)=\widehat{m}_t(\theta_0,T)$ for $t=1,\ldots,\theta_0$. One can see that
    \begin{align}\label{ineq::thm component estimator ess2}
        &{\rm pr}\left( \max_{t=1,\ldots,T} d(\widehat{m}(t),m(t)) >\frac{C}{T^{-1/2(\beta^\prime-1)} } , \widehat{\theta}_{\widehat{\lambda}} = \theta_0\right) \nonumber\\
        =& {\rm pr}\left( \max_{t=1,\ldots,\theta_0} d(\widehat{m}(t),m(t)) >\frac{C}{T^{-1/2(\beta^\prime-1)} } , \widehat{\theta}_{\widehat{\lambda}} = \theta_0\right) \nonumber\\
        \leq & \sum_{t=1}^{\theta_0} {\rm pr} \left( d(\widehat{m}(t),m(t)) >\frac{C}{T^{-1/2(\beta^\prime-1)} } \right) \nonumber\\
        \leq & \sum_{t=1}^{\theta_0}  {\rm pr}\left( d(\widehat{m}(t),\widetilde{m}(t)) >\frac{C}{T^{-1/2(\beta^\prime-1)} } \right) + \sum_{t=1}^{\theta_0}  {\rm pr}\left( d(\widetilde{m}(t),m(t)) >\frac{C}{T^{-1/2(\beta^\prime-1)} } \right).
    \end{align}
    Observe that
    \begin{align}\label{ineq::thm component estimator ess3}
        {\rm pr}\left( d(\widehat{m}(t),\widetilde{m}(t)) >\frac{C}{T^{-1/2(\beta^\prime-1)} } \right) \leq &{\rm pr}\left( d(\widehat{m}(t),\widetilde{m}(t)) >\frac{C}{T^{-1/2(\beta^\prime-1)} } ,\widehat{\theta}_{\widehat{\lambda}} = \theta_0 \right) \nonumber \\
        &+ {\rm pr}(\widehat{\theta}_{\widehat{\lambda}} \neq \theta_0).
    \end{align}
    First note that $d(\widehat{m}(t),\widetilde{m}(l)) =0$ when $\widehat{\theta}_{\widehat{\lambda}} = \theta_0$. This means 
    \[
    {\rm pr}\left( d(\widehat{m}(t),\widetilde{m}(t)) >\frac{C}{T^{-1/2(\beta^\prime-1)} } ,\widehat{\theta}_{\widehat{\lambda}} = \theta_0 \right) = 0,
    \]
    which, together with (\ref{ineq::thm component estimator ess1}), ensures that the right side of (\ref{ineq::thm component estimator ess3}) is $o(1)$.

    By the convergence rate given in Proposition~\ref{prop::uniform convergence rate partial mean}, we have the pointwise convergence rate $d(\widetilde{m}(t),m(t))=O_p(T^{-1/2(\beta^\prime-1)})$ holds for any $t=1,\ldots,\theta_0$. This result, in conjunction with (\ref{ineq::thm component estimator upper bound}), (\ref{ineq::thm component estimator ess2}) and (\ref{ineq::thm component estimator ess3}), implies that (\ref{ineq::thm component estimator goal}). This completes the proof.
    \end{proof}

\section{Additional Figure on Periodic Spherical Data}\label{sub sect::simulation sphere additional}

\begin{figure}[!htbp]
\centering
\includegraphics[width=1\linewidth]{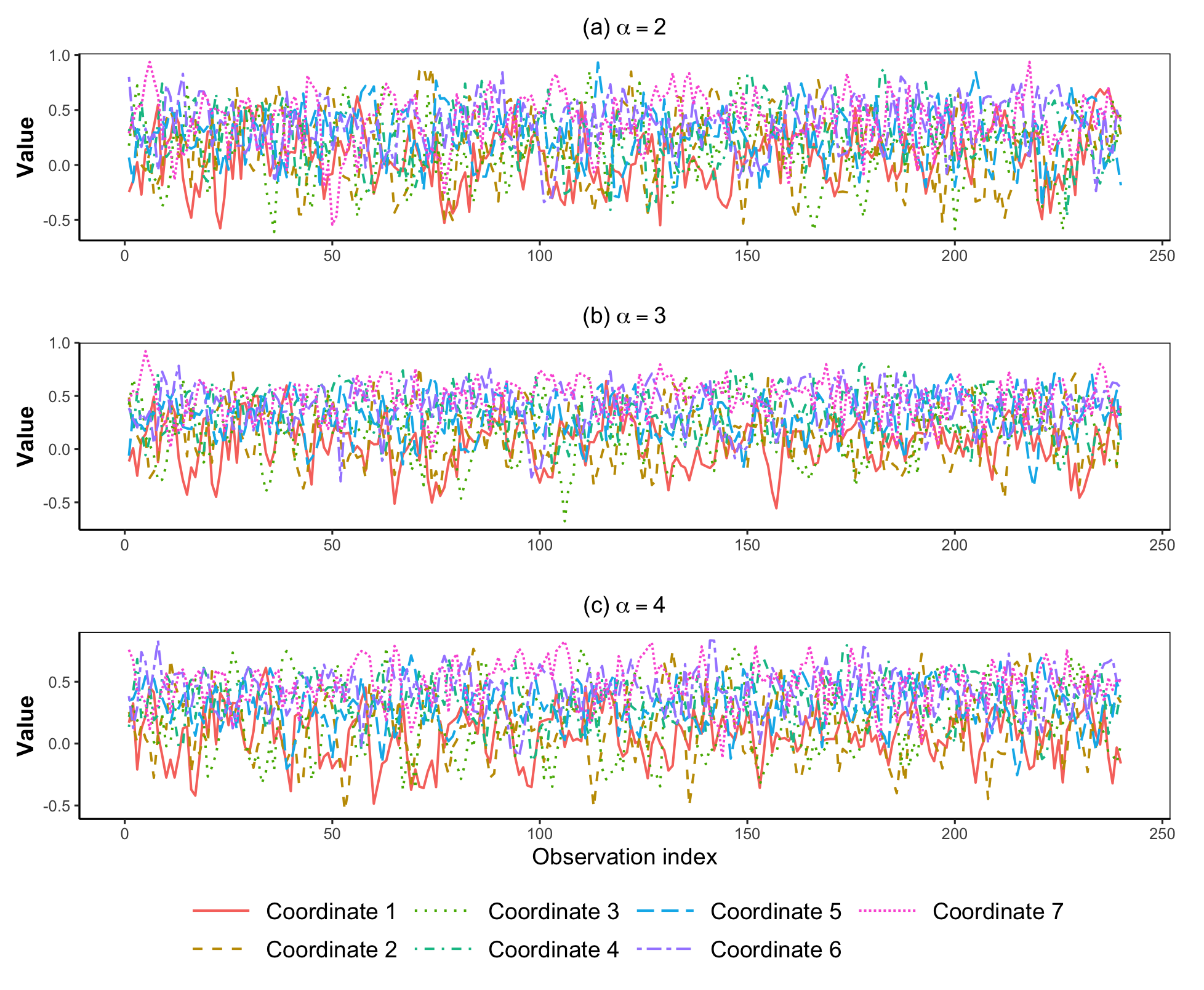}

\caption{Plots of the generated periodic spherical time series with difference random noise level (a) high noise level where $\alpha=2$, (b) median noise level where $\alpha=3$, and (c) low noise where level $\alpha=4$ \label{fig::periodic spherical ts}}
\end{figure}

Figure~\ref{fig::periodic spherical ts} visualizes the generated periodic spherical time series with $T=240$ under three difference noise levels: (a) high noise level when $\alpha=2$, (b) median noise level when $\alpha=3$, and (c) low noise level $\alpha=4$. 

\section{Additional Simulation Studies on Periodic Networks}\label{sub sect::simulation network}

In the remaining simulation studies for networks and functions, we mainly focus on period estimation for $\theta_0$, which is the first and crucial step in our proposed estimation methodology. The performance of the periodic component estimator $\widehat{m}(t)$ for $t=1,\ldots,T$ depends on the finite-sample performance of the period estimator $\widehat{\theta}_{\widehat{\lambda}}$. If the period $\theta_0$ is known, the periodic component estimator $\widehat{m}(t)$ simplifies to the global Fr\'{e}chet regression estimator which has been well studied in \cite{petersen2019frechet} and \cite{zhou2022network}. The finite sample performance of the  the periodic component estimator $\widehat{m}(t)$ is fully studied in \S\ref{sub sect::simulation composition}. Given a good estimator of $\theta_0$, the estimator $\widehat{m}(t)$ can be expected to perform similarly to the standard estimator in global Fr\'{e}chet regression for $t=1,\ldots,T$. For this reason, we mainly focus on evaluating the properties of $\widehat{\theta}_{\widehat{\lambda}}$ in numerical studies for periodic networks and functions.

\subsection{Periodic Network Generation}\label{supp sect:: network generate}

To generate networks with periodic behaviour, we use a time-varying network generation techniques similar to those in \cite{dubey2021modeling} and \cite{xu2025change}. The number of nodes in the simulated time-varying networks is 20 and we index the nodes of the networks by $1, 2,\ldots,20$ and the five communities of the nodes in this setup are given by $G_1=\{1, 2, 3, 4\},G_2=\{5, 6, 7, 8\},G_3=\{9, 10, 11, 12\},G_4=\{13, 14, 15, 16\}$ and $G_5=\{17, 18, 19, 20\}$.

\begin{figure}[!htbp]
\centering
\includegraphics[width=1\linewidth]{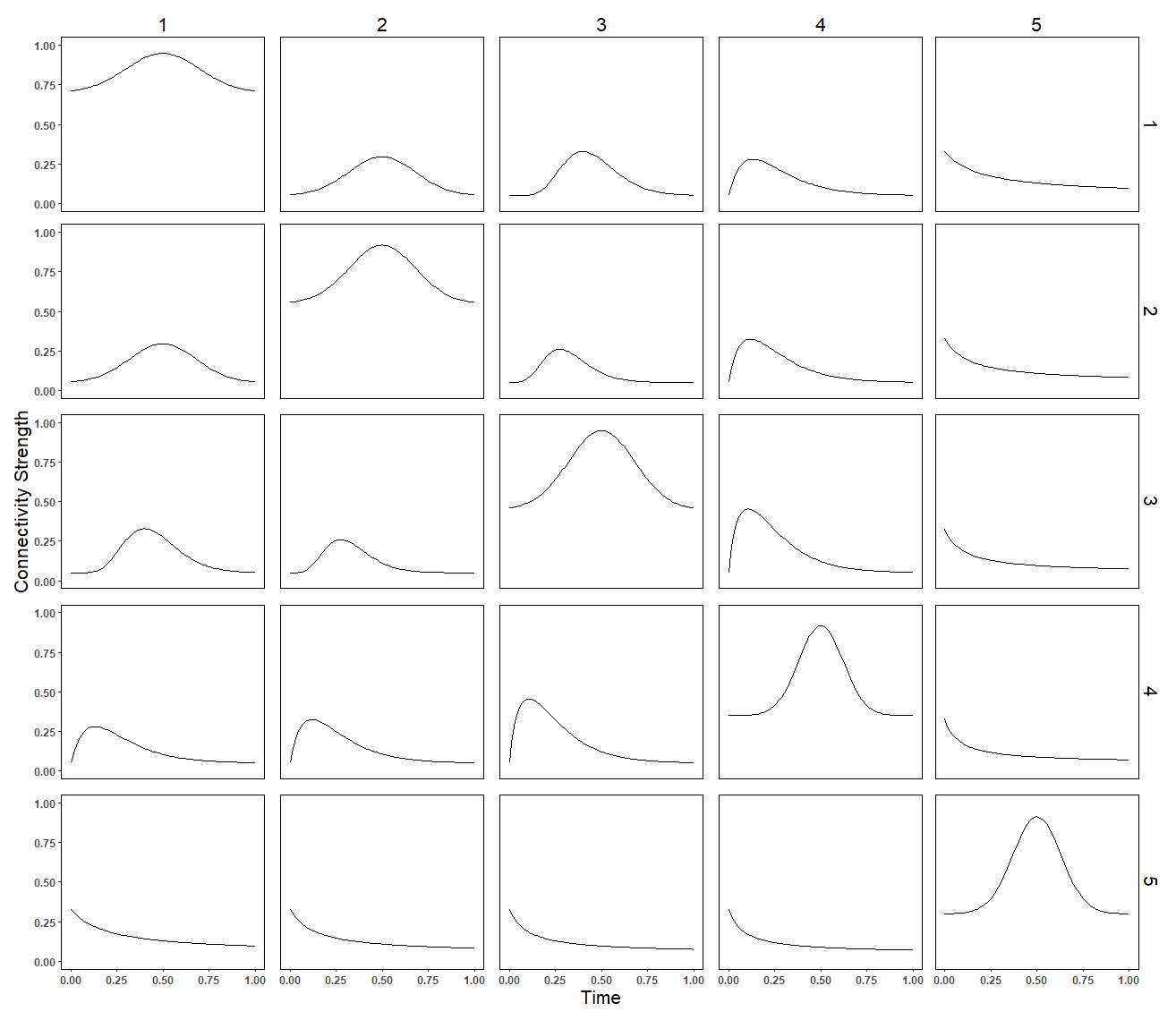}

\caption{Time varying connectivity weights between the five communities $C_1,\ldots,C_5$ for uncontaminated data.\label{fig::connectivity_weight}}
\end{figure}

Community memberships were fixed over time, while edge connectivity strengths between communities varied over time. The weights between communities $G_j$ and $G_{j^\prime}$ ($j,j^\prime\in\{1, 2, 3, 4, 5\}$), denoted as $W_{jj^\prime}(t)$, used in generating the random networks, are shown in Fig.~\ref{fig::connectivity_weight}.

Let $G_j$ be the community membership of node $j$ and $G_{j^\prime}$ be the community membership of node $j^\prime$.  
The network adjacency matrices $A_k(t)$ are generated as follows for $k\in\{1,2,\ldots,K\}$:
\begin{align}\label{eq::adj matrix generate simulation}
(A_k(t))_{j,j^\prime}=W_{j,j^\prime}(t)\left\{ \frac{\sin ((t+R_{k,j,j^\prime})S_{k,j,j^\prime}\pi)+1}{D} \right\}+ |\epsilon|_{j,j^\prime,t} ,~t\in[0,1],
\end{align}
where $\epsilon_{j,j^\prime,t}=\epsilon_{j^\prime,j,t}$ controls the additional contaminated noise and is generated independently and identically distributed from a normal distribution $N(0,\sigma^2_\epsilon)$, $W_{jj^\prime}(t)$ is the edge connectivity weight function between nodes in communities $G_j$ and $G_{j^\prime}$, $R_{k,j,j^\prime}$ is sampled from the uniform distribution $U(0, 1)$ and $S_{k,j,j^\prime}$ is set to 1 if $j=j^\prime$ or sampled uniformly from $\{5,6,\ldots,15\}$ otherwise. The parameter $D$ is set to 2 for $j=j^\prime$ and 4 otherwise. One can see that $R_{k,j,j^\prime}$ and $S_{k,j,j^\prime}$ control random phase and frequency shifts of the sine function, affecting the timing and frequency of zero weights. Higher values of $S_{k,j,j^\prime}$ increase the frequency of zero weights within $[0, 1]$. The time-varying graph Laplacians are generated by $Z_k(t)=D_k(t)-A_k(t)$, with $D_k(t)$ as a diagonal matrix whose diagonal elements are row sums of $A_k(t)$.

To obtain $\theta_0$-periodic networks with sample size $T$, we first generate $K=\lfloor T/\theta_0 \rfloor$ periodic blocks such that $\bm{Z}_j=(Z_j(t_1),Z_j(t_2),\ldots,Z_j(t_{\theta_0}))$ for $j\in\{1,2,\ldots,K\}$ and $0\leq t_1 < t_2 <\cdots<t_{\theta_0}\leq 1$. Note that the observations $Y_1,Y_2,\ldots,Y_T$ correspond to the first $T$ observations from $\bm{Z}_1,\bm{Z}_2,\ldots,\bm{Z}_K$. In the following simulations for periodic networks, we consider $\theta_0=24$.

\begin{figure}[!htbp]
\centering
\includegraphics[width=1\linewidth]{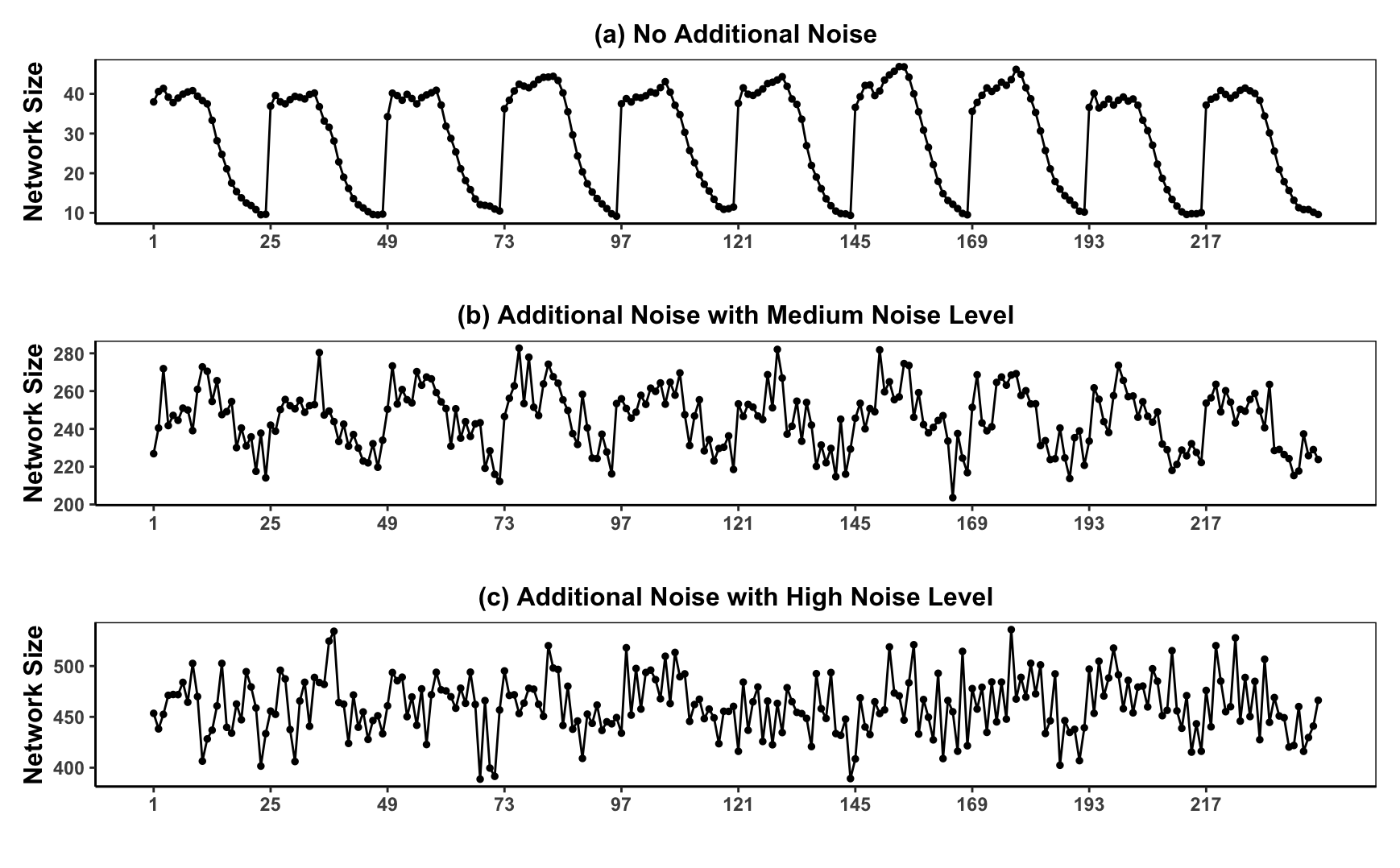}

\caption{ Plot of network sizes of the simulated networks (a) without additional noise, (b) with additional noise and the noise level is medium, and (c) with additional noise and the noise level is high.\label{fig::simulated network}}
\end{figure}

To illustrate the periodic behaviour of the simulated networks and the effect of the noise level $\sigma^2_\epsilon$ in (\ref{eq::adj matrix generate simulation}), we calculate the network size by summing all elements of the corresponding network adjacency matrix. Fig.~\ref{fig::simulated network} presents network sizes of simulated 240 networks with $\theta_0=24$ and $\sigma^2_\epsilon$ varying in $\{0,1,4\}$.

Notably, when $\sigma^2_\epsilon=0$, we do not have any additional noise and the randomness is coming from the network generation as seen in the first part of the right side in (\ref{eq::adj matrix generate simulation}). When $\sigma^2_\epsilon=1$, the noise level is medium, compared to the range of the uncontaminated adjacency matrix elements whose range is from 0 to 3.  When $\sigma^2_\epsilon=4$, the noise level is high. The clear periodic pattern in network sizes, evident in Fig.~\ref{fig::simulated network} with no additional noise. When the noise level increases, the periodic pattern becomes harder to be directly observed via the network sizes in Fig.~\ref{fig::simulated network}.

\subsection{Simulation Results}

The metric we consider for the network Laplacians in this simulation is the Frobenius metric. To fit a global Fr\'{e}chet regression for network Laplacians, we follow the techniques proposed by \cite{zhou2022network} and the corresponding R code is available at \url{https://github.com/yidongzhou/Network-Regression-with-Graph-Laplacians}. The sample size $T$ varies in $\{240,500\}$ while the upper bound of the period candidates $\Theta_T$ grows at the rate of $T^{1/3}$ and varies in $\{111,142\}$, respectively.

\begin{table}[!htbp]
\caption{Empirical probabilities that $\widehat{\theta}_{\widehat{\lambda}}=24$, reported in the first two columns, and that $20 \leq \widehat{\theta}_{\widehat{\lambda}}\leq 28$, reported in the last two columns, for the sample size $T$ varies in $\{240,500\}$ and different additional noise levels when the simulated samples are periodic networks and the information criterion is for the logarithm of residual sum of squares. \label{table::network simulation period est criteria 1}}
\centering
\scalebox{0.9}{
\begin{tabular}{l | cc | cc }
  \hline
&\multicolumn{2}{c|}{$p(\widehat{\theta}_{\widehat{\lambda}}=24)$}  & \multicolumn{2}{c}{$p(20 \leq \widehat{\theta}_{\widehat{\lambda}}\leq 28)$}  \\
  & $T=240$ & $T=500$  & $T=240$ & $T=500$ \\ 
  \hline
No additional noise & 1.000 & 1.000 & 1.000 & 1.000 \\ 
Medium additional noise & 0.000 & 0.940 & 0.000 & 0.940 \\ 
High additional noise & 0.000 & 0.930 & 0.000 & 0.930 \\ 
  \hline
\end{tabular}
}
\end{table}

\begin{table}[!htbp]
\caption{Empirical probabilities that $\widehat{\theta}_{\widehat{\lambda}}=24$, reported in the first two columns, and that $20 \leq \widehat{\theta}_{\widehat{\lambda}}\leq 28$, reported in the last two columns, for the sample size $T$ varies in $\{240,500\}$ and different additional noise levels when the simulated samples are periodic networks and the information criterion is for the residual sum of squares. \label{table::network simulation period est criteria 2}}
\centering
\scalebox{0.9}{
\begin{tabular}{l | cc | cc }
  \hline
&\multicolumn{2}{c|}{$p(\widehat{\theta}_{\widehat{\lambda}}=24)$}  & \multicolumn{2}{c}{$p(20 \leq \widehat{\theta}_{\widehat{\lambda}}\leq 28)$}  \\
  & $T=240$ & $T=500$  & $T=240$ & $T=500$ \\ 
  \hline
No additional noise & 1.000 & 1.000 & 1.000 & 1.000 \\ 
Medium additional noise & 0.845 & 1.000 & 0.845 & 1.000 \\ 
High additional noise & 0.000 & 0.925 & 0.000 & 0.925 \\ 
  \hline
\end{tabular}
}
\end{table}

Table \ref{table::network simulation period est criteria 1} presents the simulation results for network Laplacians using the first IC in (\ref{eq::BIC}) while Table \ref{table::network simulation period est criteria 2} shows the simulation results for the second IC in (\ref{eq::BIC}).  The first three columns displayed in Tables \ref{table::network simulation period est criteria 1} and \ref{table::network simulation period est criteria 2} report the probability that the estimator $\widehat{\theta}_{\widehat{\lambda}}$ correctly identifies the true period $\theta_0=24$, while the last three columns of the tables give the probabilities of $\widehat{\theta}_{\widehat{\lambda}}$ falling within the range between 20 and 28. Notably, our proposed estimator exhibits high accuracy even with a small sample size when no additional noise is introduced, achieving a probability of 1 when $\widehat{\theta}_{\widehat{\lambda}}=24$ for the sample size $T=240$. However, when additional noise is introduced, the estimator's performance deteriorates for small sample sizes, as the noise distorts the underlying periodic patterns. Nevertheless, as the sample size increases, the estimator’s accuracy improves significantly. Notably, Tables \ref{table::network simulation period est criteria 1} and \ref{table::network simulation period est criteria 2} show that the finite-sample performance of the second IC is better than the first IC.

\begin{figure}[!htbp]
\centering
\includegraphics[width=1\linewidth]{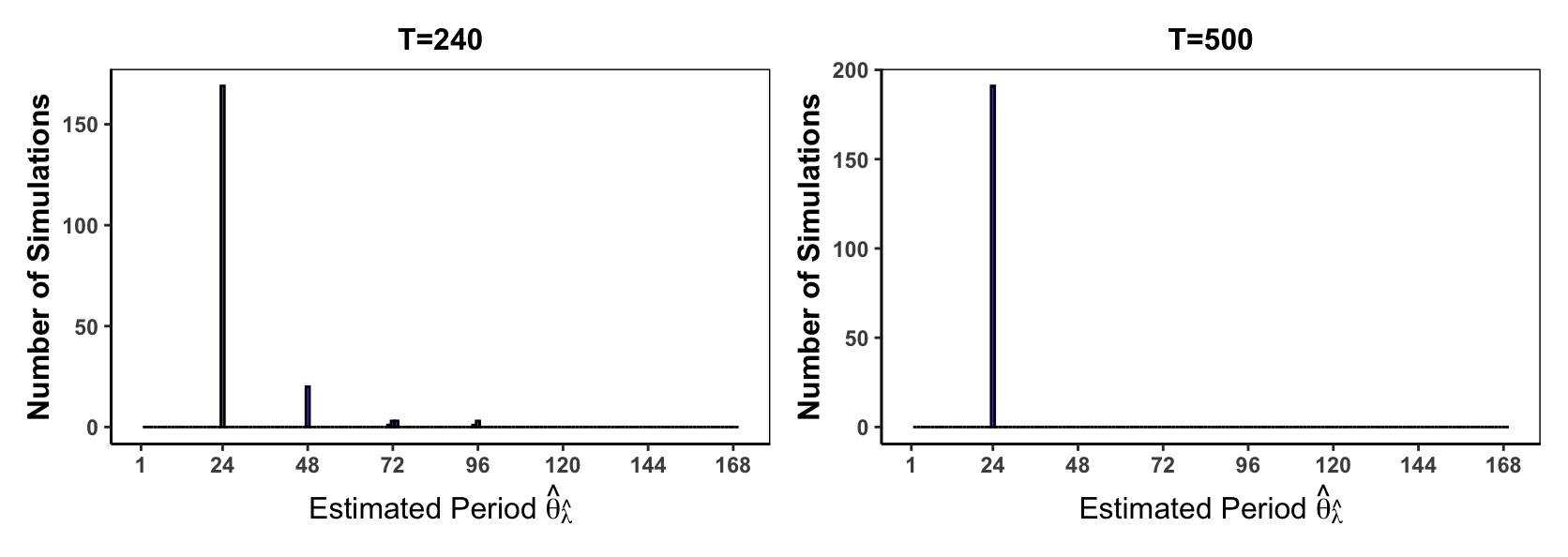}

\caption{ Histograms of the simulation results for periodic networks when the additional noise level is medium for different samples sizes $T\in\{240,500\}$. The bars present how many times each value of $\widehat{\theta}_{\widehat{\lambda}}$ is observed across 200 simulation runs.  \label{fig::network simulation period est hist}}
\end{figure}

Figure~\ref{fig::network simulation period est hist} provides a graphical summary of the simulation results under medium noise levels for different sample sizes $T\in\{240,500\}$. Each panel illustrates the behaviour of $\widehat{\theta}_{\widehat{\lambda}}$ for the respective sample sizes. Similar to Fig.~\ref{fig::composition simulation period est hist},  alongside the primary cluster of estimates near the true period $\theta_0$, smaller clusters also appear around its integer multiples. Although Table \ref{table::network simulation period est criteria 1} suggests that the estimator may not always precisely recover the true period with medium noise level, it still provides reasonable approximations, particularly capturing its multiples. As the sample size increases, the clusters around the multiples of $\theta_0$ disappear, leading to more accurate period identification.

\begin{figure}[!htbp]
\centering
\includegraphics[width=1\linewidth]{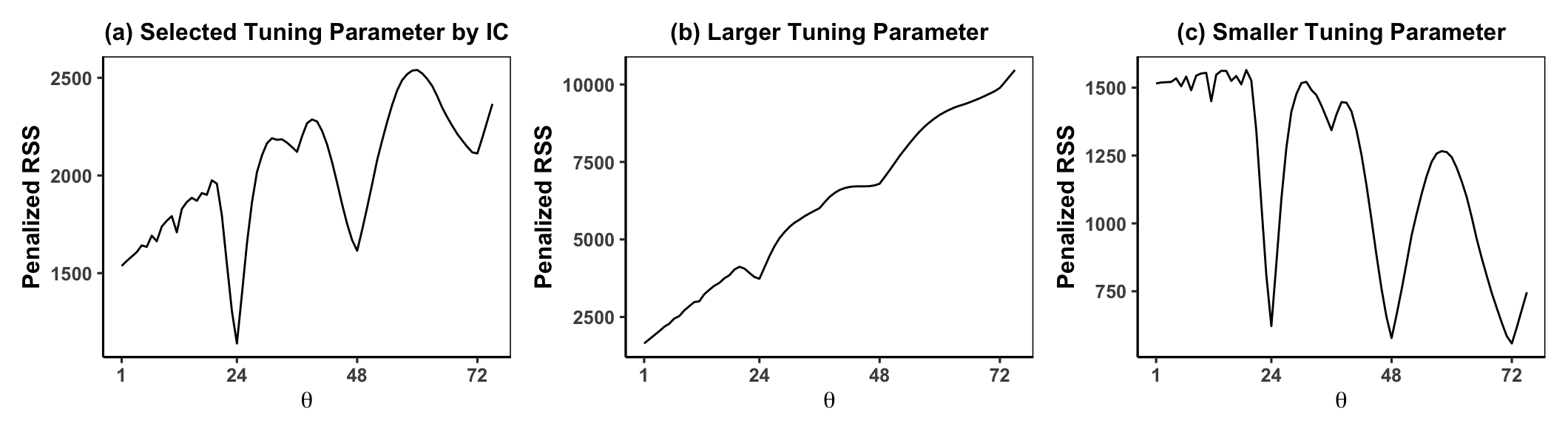}

\caption{ Plots of penalized RSS for one simulation replicate for periodic networks with $T=100$ and no additional noise considering different choices of the tuning parameter $\lambda_T$ such as $\lambda_T=\widehat{\lambda}$ in (a), $\lambda_T=5\widehat{\lambda}$ in (b), and $\lambda_T=\widehat{\lambda}/5$ in (c). \label{fig::simulation network loss}}
\end{figure}

Finally, we assess the effectiveness of our selected tuning parameter based on the proposed IC in (\ref{eq::BIC}). We conduct a typical simulation for periodic networks with $T=100$ and no additional noise. Similar to Fig.~\ref{fig::simulation composition loss}, Fig.~\ref{fig::simulation network loss} presents the penalized RSS $\mathcal{L}(\theta,\lambda_T)$ from (\ref{formula::loss function}) across three different settings of the tuning parameter $\lambda_T$. In panel (a), $\lambda_T=\widehat{\lambda}$ corresponds to the value selected via the proposed IC. In panel (b), we choose a larger tuning parameter, $\lambda_T=5\widehat{\lambda}$ while in panel (c), we use a smaller tuning parameter as $\lambda_T=\widehat{\lambda}/5$. Across both panels in Fig.~\ref{fig::simulation network loss} when the tuning parameter is not too large, the penalized RSS curves show sharp drops near the true period $\theta_0=24$ and its multiples. Again, the choice of $\lambda_T$ significantly affects the function’s overall trend.

As seen in panel (b) of Fig.~\ref{fig::simulation network loss}, when the tuning parameter is large, the penalized RSS rises sharply, causing the global minimum to shift from the first downward spike at the true period to $\theta=1$. Conversely, in panel (c), a smaller tuning parameter leads to a decreasing trend, with the global minimum occurring at a later spike instead of at the true period. These findings confirm that our selected tuning parameter, based on the proposed IC, provides a reasonable and effective choice for period estimation.

\section{Additional Simulation Studies on Periodic Distributions}\label{sub sect::simulation function}

The generation of periodic functions is as follow for $t=1,\ldots,T$, $Y_t=f(x;c_t)+\epsilon_t$ where$f(x;c)=10e^{-(x-c)^2}+5$, $c_t = \sin(2\pi t/\theta_0)$ with the true period $\theta_0=7$. The error term $\epsilon_t(x)$ is constructed to exhibit both spatial smoothness and temporal correlation via a functional autoregressive process of order one, expressed as $\epsilon_t(x) = \phi \epsilon_{t-1}(x) + e_t(x)$, where $\phi=0.5$ dictates the degree of temporal dependence. The innovations $e_t(x)$ are drawn independently from a zero-mean Gaussian process governed by a squared-exponential covariance function, $\text{Cov}(e_t(x), e_t(x^\prime)) = \sigma_\epsilon^2 \exp\left\{-(x-x^\prime)^2/(2l^2)\right\}$. Here, $\sigma_\epsilon^2$ scales the noise variance and $l$ is the bandwidth controlling the smoothness of the generated curves. Finally, $Y_t$ is normalized to be the density function and the metric we consider here is the Wasserstein metric between two densities $f_1$ and $f_2$ as $d_w(f_1,f_2)=\sqrt{\int_0^1 \{Q_1(t)-Q_2(t)dt\}^2}$ where $Q_1(t)$ and $Q_2(t)$ are the corresponding quantile functions. The sample size $T$ varies in $\{100,240,500\}$ and $\Theta_T$ grows at the rate of $T^{1/3}$ and varies in $\{23,31,40\}$, respectively.

\begin{table}[!htbp]
\caption{Empirical probabilities that $\widehat{\theta}_{\widehat{\lambda}}=7$, reported in the first three columns, and that $4 \leq \widehat{\theta}_{\widehat{\lambda}}\leq 10$, reported in the last three columns, for the sample size $T$ varies in $\{100,240,500\}$ and different levels of element-wise variance of each observation when the simulated samples are periodic functions and the information criterion is for the logarithm of residual sum of squares.  \label{table::function simulation period est criteria 1}}
\centering
\scalebox{0.9}{
\begin{tabular}{l | ccc | ccc }
  \hline
&\multicolumn{3}{c|}{$p(\widehat{\theta}_{\widehat{\lambda}}=7)$}  & \multicolumn{3}{c}{$p(4 \leq \widehat{\theta}_{\widehat{\lambda}}\leq 10)$}  \\
 & $T=100$ & $T=240$ & $T=500$ & $T=100$ & $T=240$ & $T=500$ \\ 
  \hline
$\sigma_\epsilon^2 = 0.01$ & 1.000 & 1.000 & 1.000 & 1.000 & 1.000 & 1.000 \\ 
  $\sigma_\epsilon^2= 0.25$ & 1.000 & 1.000 & 1.000 & 1.000 & 1.000 & 1.000 \\ 
  $\sigma_\epsilon^2= 1$ & 1.000 & 1.000 & 1.000 & 1.000 & 1.000 & 1.000 \\ 
    \hline
\end{tabular}
}
\end{table}

\begin{table}[!htbp]
\caption{Empirical probabilities that $\widehat{\theta}_{\widehat{\lambda}}=7$, reported in the first three columns, and that $4 \leq \widehat{\theta}_{\widehat{\lambda}}\leq 10$, reported in the last three columns, for the sample size $T$ varies in $\{100,240,500\}$ and different levels of element-wise variance of each observation when the simulated samples are periodic functions and the information criterion is for the residual sum of squares. \label{table::function simulation period est criteria 2}}
\centering
\scalebox{0.9}{
\begin{tabular}{l | ccc | ccc }
  \hline
&\multicolumn{3}{c|}{$p(\widehat{\theta}_{\widehat{\lambda}}=7)$}  & \multicolumn{3}{c}{$p(4 \leq \widehat{\theta}_{\widehat{\lambda}}\leq 10)$}  \\
 & $T=100$ & $T=240$ & $T=500$ & $T=100$ & $T=240$ & $T=500$ \\ 
  \hline
$\sigma_\epsilon^2 = 0.01$ & 1.000 & 1.000 & 1.000 & 1.000 & 1.000 & 1.000 \\ 
  $\sigma_\epsilon^2= 0.25$ & 1.000 & 1.000 & 1.000 & 1.000 & 1.000 & 1.000 \\ 
  $\sigma_\epsilon^2= 1$ & 1.000 & 1.000 & 1.000 & 1.000 & 1.000 & 1.000 \\ 
    \hline
\end{tabular}
}
\end{table}

Tables \ref{table::function simulation period est criteria 1} and \ref{table::function simulation period est criteria 2} summarizes the simulation results for periodic functions. The first three columns display the probability that the estimator $\widehat{\theta}_{\widehat{\lambda}}$ correctly identifies the true period $\theta_0=7$, whereas the last three columns show the probabilities of $\widehat{\theta}_{\widehat{\lambda}}$ lying between 4 and 10. Both Tables \ref{table::function simulation period est criteria 1} and \ref{table::function simulation period est criteria 2} show that the estimator identifies the true period for any replicate, even in noisier settings.

\begin{figure}[!htbp]
\centering
\includegraphics[width=1\linewidth]{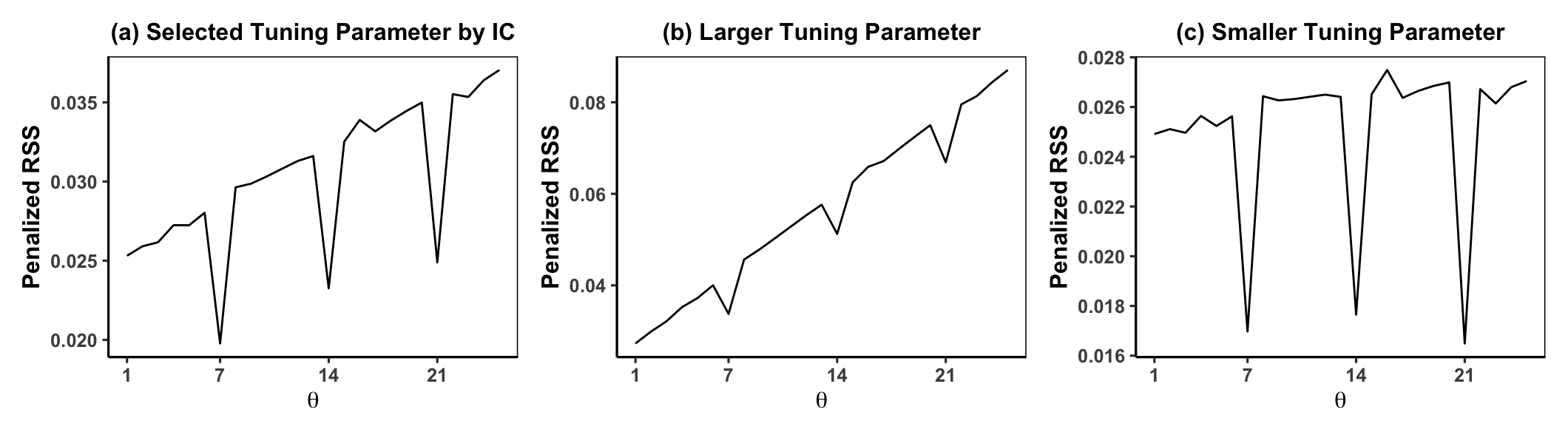}

\caption{ Plots of penalized RSS for  one simulation replicate  for periodic functions with $T=100$ and $\sigma_\epsilon^2 = 0.01$ considering different choices of the tuning parameter $\lambda_T$ such as $\lambda_T=\widehat{\lambda}$ in (a), $\lambda_T=5\widehat{\lambda}$ in (b), and $\lambda_T=\widehat{\lambda}/5$ in (c). \label{fig::simulation function loss}}
\end{figure}

Finally, we evaluate the performance of our selected tuning parameter based on the proposed IC in (\ref{eq::BIC}). We consider a typical simulation for periodic compositions with $T=100$ and $\sigma_\epsilon^2 = 0.01$ . Similar to Figs.~\ref{fig::simulation composition loss} and \ref{fig::simulation network loss}, Fig.~\ref{fig::simulation function loss} displays the penalized RSS $\mathcal{L}(\theta,\lambda_T)$ from (\ref{formula::loss function}) for three tuning parameter choices: panel (a) corresponds to $\lambda_T=\widehat{\lambda}$, panel (b) considers a larger tuning parameter $\lambda_T=5\widehat{\lambda}$, and panel (c) explores a smaller tuning parameter $\lambda_T=\widehat{\lambda}/5$. Similar to Figs.~\ref{fig::simulation network loss} and \ref{fig::simulation composition loss}, the true period $\theta_0=7$ and its multiples are marked by noticeable dips in the penalized RSS functions across all panels. The choice of $\lambda_T$ has a significant impact on the function’s overall trend: a larger tuning parameter, as seen in panel (b), forces the penalized RSS to increase quickly and thus the global minimum occurs at $\theta=1$. A smaller tuning parameter, as shown in panel (c), shifts the global minimum to a later spike instead of the true period. These results further validate that our selected tuning parameter, based on the proposed IC, provides a reasonable and effective choice.

\section{Additional Figure for Real Data Analysis}\label{supp sect:: real data figure}

\begin{figure}[!htbp]
\centering
\includegraphics[width=1\linewidth]{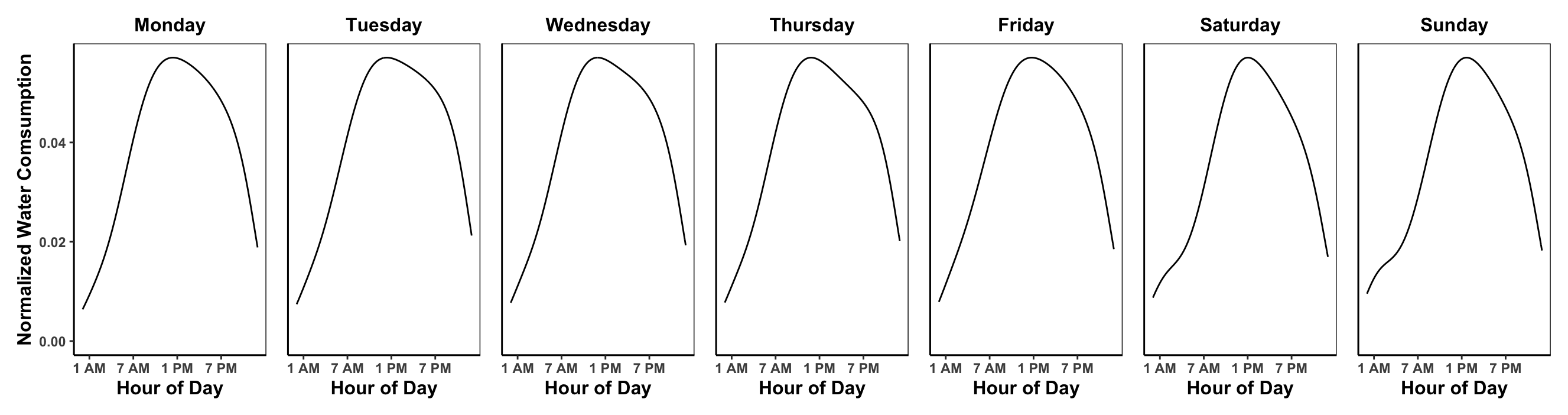}

\caption{Plots of the estimated periodic component of the normalized daily water consumption curves in Germany. The estimated periodic component has a period 7 and contains 7 curves. \label{fig::real data period function component est}}
\end{figure}

Figure~\ref{fig::real data period function component est} shows  all the elements in the estimated periodic component. As mentioned in \S\ref{subsect::water application}, the daily water consumption curves have similar shapes on weekdays and weekends, respectively. This leads to a period of 7 days and a intra-week periodic pattern in the water consumption curves.

\clearpage
\newpage

\bibliographystyle{biometrika}